# The Hitchhiker's Guide to Programming and Optimizing Cache Coherent Heterogeneous Systems
## *CXL, NVLink-C2C, and AMD Infinity Fabric*

Zixuan Wang[Ψ], Suyash Mahar[Ψ], Luyi Li[Ψ], Jangseon Park[Ψ✸],
Jinpyo Kim[Ψ✱], Theodore Michailidis[Ψ], Yue Pan[Ψ], Mingyao Shen[Ψ],
Tajana Rosing[Ψ], Dean Tullsen[Ψ], Steven Swanson[Ψ], and Jishen Zhao[Ψ]

[Ψ]University of California, San Diego  [✸]Samsung  [✱]SK Hynix

**Abstract**

We present a thorough analysis of the use of modern heterogeneous systems interconnected by various cache-coherent links, including CXL, NVLink-C2C, and Infinity Fabric. We studied a wide range of server systems that combined CPUs from different vendors and various types of coherent memory devices, including CXL memory expander, CXL pool, CXL shared memory, GH200 GPU, and AMD MI300a HBM. For this study, we developed a heterogeneous memory benchmark suite, HEIMDALL, to profile the performance of such heterogeneous systems and present a detailed performance comparison across systems. By leveraging HEIMDALL, we unveiled the detailed architecture design in these systems, drew observations on optimizing performance for workloads, and pointed out directions for future development of cache coherent heterogeneous systems.

## 1 Introduction

The ever-growing performance demands from modern applications drive the development of heterogeneous systems. However, heterogeneous systems' communication bandwidth has become one of the key bottlenecks in system scalability, where the hardware bandwidth does not scale as fast as modern workloads' bandwidth requirements. To improve the bandwidth, devices such as network cards and GPUs have developed dedicated communication links [1] to exchange data more efficiently. However, such new communication links often define their own communication protocols, requiring specialized operating system kernel drivers and system software libraries and imposing new programming models on workload developers. Adding such new devices to an existing heterogeneous system is thus non-trivial, which can hold back their adoption. This motivates the need for an industry-standard communication protocol to provide a consistent interface to workloads while enabling device manufacturers to integrate new devices into the ecosystem without changing the programming interface.

Cache coherent interconnect protocols are proposed to unify the communication interface between heterogeneous devices. With cache-coherent interconnects, processors can access device memory through the cache-coherent bus and cache such data locally; At the same time, the protocol transparently updates the locally cached data when modified by any connected device.Such protocols allow processors to exchange data in a standard scheme, simplifying data synchronization and reducing the use of dedicated communication drivers and libraries for data synchronization. Many existing cache-coherent protocols are initially deployed for homogeneous systems such as inter-CPU links [2, 3]. With the rapid emergence of new accelerators and memory devices, industry and academia have been exploring generic cache-coherent links [4–7] for heterogeneous systems to interconnect different types of processors and memory. Such generic protocols aim to unify communication schemes and optimize data exchange performance between devices.

Compute Express Link [4] (CXL) is a recent open standard of cache coherent interconnect protocol and has been commercially supported in its early stage. CXL defined a multi-device coherence protocol on top of the PCIe physical layer, allowing processors to reuse the existing PCIe standard (links, form factors, and more) as much as possible instead of adopting a new physical layer design. When interconnected with CXL, accelerators and memory devices can sit on the PCIe bus and exchange data coherently with other devices, including the host CPU. As of today, CXL has three generations of standard specs, from 1.0 to 3.0. The CXL 1.0 spec lays the foundation of the CXL standard, defining the basics of coherence and device data exchange protocols; the revised 1.1 spec added the specification for memory expander devices. The CXL 2.0 introduced switch-based topologies, enabling scalable multi-device configurations and improved memory pooling. The recent CXL 3.0 doubled the bandwidth with PCIe Gen 6 and incorporated advanced features such as atomic operations and enhanced security, supporting more demanding applications like AI and large-scale memory pools.

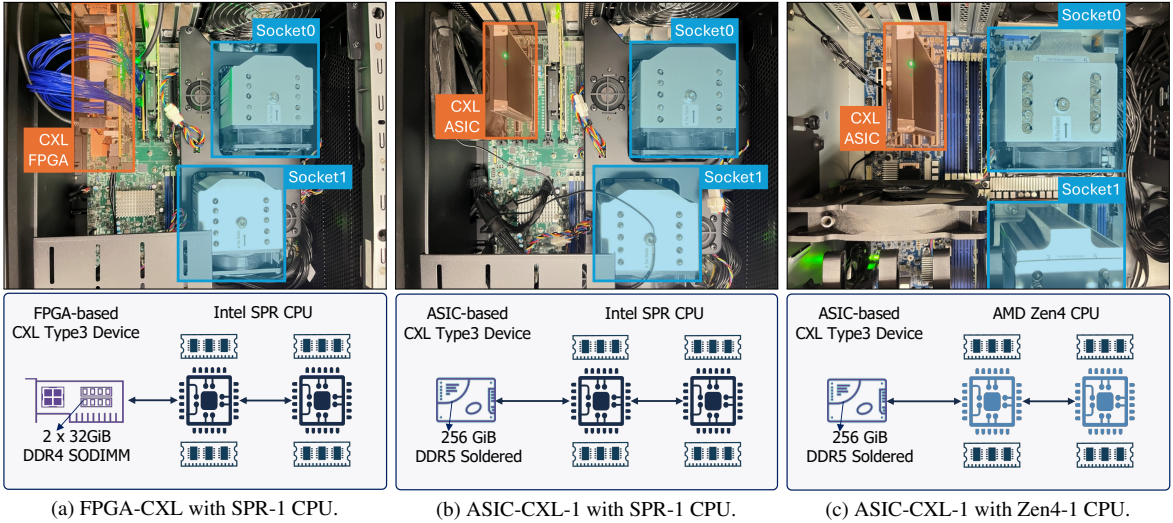

(a) FPGA-CXL with SPR-1 CPU.  (b) ASIC-CXL-1 with SPR-1 CPU.  (c) ASIC-CXL-1 with Zen4-1 CPU.

Figure 1: CXL system organization overview, the top sub-figures are photos of the real machine we've built locally, and the bottom sub-figures are illustrations of the system build. This figure shows only part of the system in our fleet, and see Table 1 for a full list of systems used in this paper.

Vendor-specific coherent protocols have also been deployed in vendors' integrated systems. NVLink-C2C [7] extends Nvidia's NVLink protocol to provide cache coherence. This protocol connects Nvidia CPUs and GPUs in Nvidia's proprietary system builds, such as GH200 and GB200, and it's not seen in other systems such as x86 CPUs. IBM Power architecture supports NVLink protocols [8], but not the C2C protocol. AMD's Infinity Fabric [3] can coherently interconnect AMD CPUs and AMD GPUs, which is commonly seen in AMD's MI300A [9] and follow-up systems.

In this paper, we then studied the performance characteristics in such heterogeneous systems, compared performance metrics side-by-side across different systems, and drew observations. To this end, we developed a benchmark suite, HEIMDALL, and leveraged it to conduct a wide range of performance profiling. This benchmark suite consists of carefully crafted microbenchmarks that trigger specific system behaviors across system layers, from microarchitecture to operating systems levels. By analyzing the benchmark result on a single system, we observed characteristics such as CPU and CXL device microarchitecture designs that support the CXL protocol, together with OS and system software performance, while leveraging CXL devices. Then by comparing benchmark results across systems, we observed discrepancies between systems, including different CPU-side CXL designs between AMD and Intel and device-side architectural implications for performance.

In summary, we make the following contributions:

- We built a cluster of CXL-based systems and summarized our lessons learned throughout this process.
- We developed a benchmark suite–HEIMDALL–for heterogeneous memory systems.
- By leveraging this benchmark suite, we studied a wide range of cache-coherent heterogeneous system configurations.
- We draw key observations from our extensive experiments and point out strategies in optimizing system software for cache-coherent heterogeneous systems.

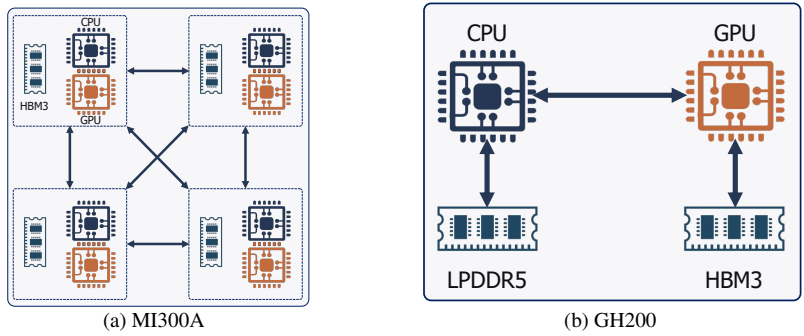

Figure 2: Topology for AMD MI300A with Infinity Fabric, and GH200 with NVLink.

## 2 Background and Methodology

### 2.1 Compute Express Link (CXL)

Compute Express Link (CXL) [4] is an open industry-standard interconnection protocol designed to facilitate high-speed communications between processors, accelerators (such as GPUs and SmartNICs), and memory devices in computing systems. A unique feature of CXL is its capability at the hardware level to enable coherent memory access across heterogeneous architectures. This coherence ensures that various processors and accelerators can transparently share memory resources as though they were part of a unified system. By leveraging CXL, we can simplify the memory access operations in heterogeneous programming models.

#### 2.1.1 CXL Protocol and Device Types

CXL supports three different protocols: CXL.io, CXL.cache, and CXL.mem. CXL.io is identical to the PCIe protocol and enables device discovery, enumeration, and PCIe IO transactions. On the other hand, CXL.cache and CXL.mem enable coherent cache communication among devices. CXL.mem allows the host to coherently access device memory, and CXL.cache additionally allows the device to coherently cache host memory on the device.

CXL defines three device types, Type 1 to 3, for different use cases: Type 1 devices are traditional PCIe devices and support only the CXL.cache protocol. Type 2 devices are accelerators with onboard memory (e.g., GPUs) and support all three CXL protocols. Finally, Type 3 devices have onboard memory for capacity and bandwidth expansion but do not cache host memory; they support CXL.io and CXL.mem.

The CXL systems we tested (Figure 1 and Table 1) provide three types of interfaces to access the CXL on-device resources: (1) All of our CXL devices expose PCIe config space and user-defined memory-mapped registers for configuring CXL device behaviors. Type 1 and 2 devices expose an additional PCIe interface for programs to interact with on-device accelerators. (2) Type 2 and 3 devices enumerate their on-device memory as an additional NUMA memory node for programs to access such memory. This ensures any existing workload can easily use CXL memory through the NUMA interface instead of specialized libraries. (3) Selectively, type 2 and 3 devices can be configured to expose their memory as a Direct Access (DAX) file–a memory-mapped device file in Linux systems–allowing workloads to use `mmap()` to access CXL memory. Once enabled, this DAX mode removes the corresponding amount of memory from the NUMA memory node to prevent conflict access from both sides.

#### 2.1.2 CXL Enumeration

To boot a system with CXL devices, CPU firmware and operating system kernel have to support CXL. CXL device enumeration is the process where the BIOS/UEFI and OS discover and identify devices connected to the CXL bus. This process is required to establish communication between the host CPU and CXL devices.

During the early bootup stage, the BIOS/UEFI automatically probes all devices connected to the PCIe (including CXL devices) and establishes early communication with CXL devices. Different CPUs handle this process differently. For example, our SPR-1 CPUs can fully enumerate the Intel-FPGA-based CXL devices without the help of the OS kernel, while our AMD CPUs cannot correctly enumerate the FPGA-based devices. CPU vendors can implement dedicated CXL enumeration processes for their CXL products (which may have specialized designs such as accelerator communication protocols and device IDs), leaving the OS kernel to enumerate generic CXL devices that comply with CXL standards. We found our SPR-1 CPU using the specialized PCIe device ID (`0x0ddb`) provided by Intel CXL FPGA to identify such devices and properly enumerate them, while our AMD CPUs do not recognize such ID.

Table 1: System specifications. EMR=Emerald Rapids, SPR=Sapphire Rapids, GH=Grace Hopper

| Machine | CPU Model | Coherent Fabric | CPU Memory / Socket | Coherent Device | Device Size |
|---|---|---|---|---|---|
| EMR-1 | 2 x Intel Xeon Gold 6542Y | CXL v1.1+ | 8 x 32 GiB DDR5 | Pool-FPGA-CXL (Pool) | 512 GiB |
| EMR-2 | 2 x Intel Xeon Gold 6530 | CXL v1.1+ | 8 x 32 GiB DDR5 | ASIC-CXL-3 (Expander) | 256 GiB |
| EMR-3 | 2 x Intel Xeon Silver 4509Y | CXL v1.1+ | 4 x 32 GiB DDR5 | SHM-FPGA-CXL (Shared Mem) | 1024 GiB |
| SPR-1 | 2 x Intel Xeon Silver 4416+ | CXL v1.1 | 4 x 32 GiB DDR5 | FPGA-CXL (Expander) | 64 GiB |
| SPR-1 | 2 x Intel Xeon Silver 4416+ | CXL v1.1 | 4 x 32 GiB DDR5 | ASIC-CXL-1 (Expander) | 256 GiB |
| SPR-2 | 2 x Intel Xeon Gold 6442Y | CXL v1.1 | 8 x 128 GiB DDR5 | ASIC-CXL-2 (Expander) | 256 GiB |
| SPR-3 | 2 x Intel Xeon Gold 5416S | CXL v1.1 | 8 x 32 GiB DDR5 | ASIC-CXL-1 (Expander) | 256 GiB |
| Zen4-1 | 2 x AMD EPYC 9124 | CXL v1.1+ | 4 x 16 GiB DDR5 | ASIC-CXL-1 (Expander) | 256 GiB |
| Zen4-2 | 4 x AMD MI300A | Infinity Fabric | 8 x 16 GiB HBM3 | HBM3 (NUMA node) | 128 GiB |
| Zen4-3 | 1 x AMD EPYC 9534 | CXL v1.1+ | 2 x 32 GiB DDR5 | Pool-FPGA-CXL (2x Interleaved) | 512 GiB |
| Zen5-1 | 2 x AMD EPYC 9535 | CXL v2.0 | 4 x 64 GiB DDR5 | ASIC-CXL-3 (Expander) | 256 GiB |
| GH200 | 1 x NVIDIA Grace | NVLink C2C | 8 x 60 GiB LPDDR5x | HBM3 (GPU) | 96 GiB |

After BIOS/UEFI finishes booting, the OS kernel (the Linux kernel in our paper) takes over and relies on its CXL device driver to further set up CXL devices. This process includes discovering CXL devices from PCIe devices, extracting CXL-related information from config registers (such as CXL-attached memory size and start address), creating corresponding Linux device files, and adding a new NUMA node for CXL-attached memory, if any. The Linux kernel support for CXL is still under development and refinement, and only until Linux 6.8-rc2 does it correctly recognize Intel FPGA-based Type 3 devices [10], and more supports coming under the path.

## 2.2 Accessing CXL Resources

There are three types of CXL devices with different combinations of CXL resources. Type 1 devices are pure accelerator devices without memory exposed to host CPUs. Type 3 devices are pure memory devices without exposing any accelerator to host CPUs. Type 2 provides both accelerator and memory resources for the host.

To access CXL accelerator resources provided by Type 1 and Type 2, a program communicates with the CXL's PCIe device file provided by the Linux kernel and follows the accelerator's communication protocol. To access CXL memory resources provided by Type 2 and Type 3, a program can chose to access through NUMA interface or a direct-access (DAX) device file. CXL memory is by default exposed as an additional NUMA memory node in the Linux system, allowing any program to easily access it through NUMA libraries or the `numactl` command line tool. Selectively, users or programs can configure the device to expose CXL memory through a DAX device file through the `daxctl` command line tool. Once configured, the program can use the `mmap()` system call to map this DAX file into its address space and then directly manage this memory space. Any access through such a DAX file will directly access the CXL memory without any DRAM memory as an intermediate cache. Such a method gives the program more flexibility in managing CXL memory space [11].

## 2.3 NVLink-C2C and Infinity Fabric

We studied the GH200 system with NVLink-C2C links, and the AMD MI300A system with Infinity Fabric links. As shown in Figure 2, the GH200 system exposes the on-GPU HBM3 to the CPU as a NUMA memory node in addition to its LPDDR5 memory node, so both memory nodes can be accessed using CPU load and store instructions. On the other hand, the AMD MI300A has its HBM3 shared by each CPU and GPU pair, and it is the only NUMA memory node for each socket.

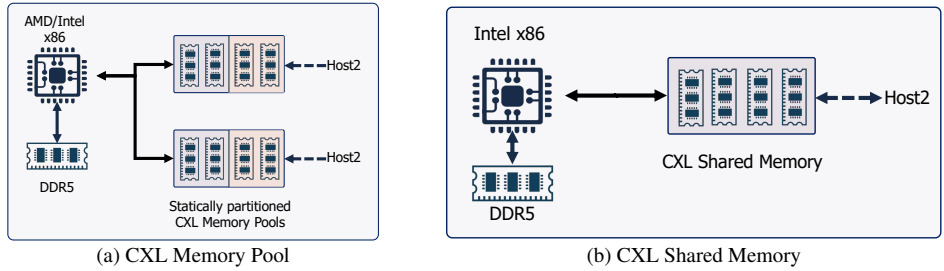

(a) CXL Memory Pool  (b) CXL Shared Memory

Figure 3: Topology for the CXL Pool (statically partitioned, unshared) and CXL Shared Memory, software managed, all of the device is accessible by both hosts. (a) and (b) use DDR4 as the device memory and are FPGA-based.

## 2.4 System Description and Experimental Configurations

Table 1 lists the systems we studied in this work, including six machines we built locally (partially depicted in Figure 1), and other machines are accessed remotely from various sources. These systems cover many of the recent generations of CPUs from Intel, AMD, and Nvidia, as well as various types of CPU memories such as DDR5, LPDDR5x, and HBM3. Note the CPUs with CXL v1.1 only supports CXL.io and CXL.cache, the v1.1+ and v2.0 additionally support CXL.mem. On the device side, we integrated three types of ASIC-based memory expanders from various vendors into Intel and AMD CPUs for evaluation. We also implemented all three types of CXL devices on Intel Agilex 7 FPGAs with 64 GiB DDR4 DRAM and evaluated them with Intel CPUs. Unless otherwise stated, we use FPGA Type 3 devices in our performance evaluations to compare with the ASIC-based memory expander. The CXL pool and CXL shared memory systems are FPGA-based prototypes hosted by our research collaborations. Their topology is shown in Section 2.4. And the Zen4-3-Pool-CXL system interleaves two CXL pool endpoints with hardware interleaving. The GH200 and MI300A systems are commercially available products that we accessed remotely.

## 2.5 HEIMDALL Framework

HEIMDALL comes with a large set of benchmarks and profiling tools to profile the system performance. This section provides an overview of these components, leaving more detailed descriptions for later sections.

**Microbenchmarks.** We develop a wide range of microbenchmarks to study the performance at the hardware and microarchitecture level. These microbenchmarks are implemented to trigger specific hardware behaviors; thus, we detect microarchitecture designs by monitoring their performance results under different configurations. As an example, we utilize our pointer-chasing microbenchmarks to study memory latency, and by varying pointer-chasing configurations, we reverse-engineer the memory architecture such as memory controller design and CXL data path. To ensure HEIMDALL microbenchmarks behave as expected, we implement their core functionalities in raw x86 assembly code to rule out potential compiler optimizations.

**Application benchmarks.** We integrate application-level benchmarks into HEIMDALL to study the system and software performance. We configure these benchmarks to run on various hardware system settings, such as remote vs. local memory access, various memory interleaving schemes, and different sub-NUMA clustering modes. By profiling such benchmarks, we observe the system performance limitations and highlight future directions for performance improvements.

**Profiling framework.** We develop HEIMDALL as a low-noise profiling framework that runs microbenchmark code in Linux kernel space to access physical memory at low noise. We utilize system configuration tools to reduce noises, including turning off CPU hardware prefetchers, disabling simultaneous multithreading (SMT), disabling interrupt handlers while running microbenchmarks, and boosting CPUs to performance mode through the CPU scaling governor. We adopt profiling tools—such as `perf` [12], AMD uProf [13], Intel PCM [14]—to collect hardware performance counters.

**Open source.** We open source HEIMDALL at `github.com/awesome-cxl/` to facilitate future research.

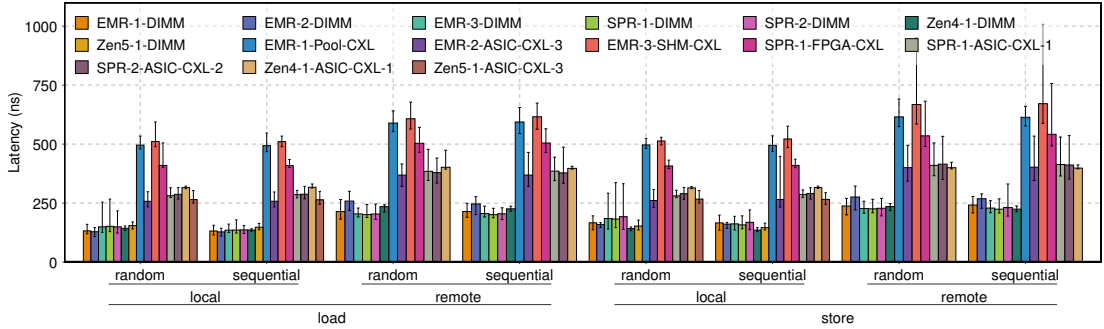

Figure 4: Load and store latency, with `clflush` after each load and store instruction to enforce access to the off-core memory. Hardware prefetchers are off. The x-axis represents the access pattern (*sequential* or *random*) and access type (accessing memory from *local* NUMA CPU or *remote* CPU), and the y-axis indicates the average access latency per cache line in nanoseconds. Each test accesses more than 32 GiB of memory and repeats 1000 times to obtain average latency values.

## 3 Basic Performance

We leverage our HEIMDALL's microbenchmark suite to study the basic performance characteristics of the CXL memory devices and compare the results with CPU-attached DDR5 DIMMs. We start with the latency and bandwidth measurements, followed by specific instructions' behaviors, including cacheline flush, prefetch, and atomic instructions. Our key observations in this measurement motivate our further investigation at the micro-architecture layer (Section 4) and application layer (Section 6).

### 3.1 Load/Store Latency

We evaluate the load and store latencies of various CXL devices and compare them against traditional DRAM DIMM. We configured HEIMDALL to accurately measure memory latency across NUMA nodes. As depicted in Figure 4, "Local" refers to memory accesses within the same NUMA node as the profiling core, whereas "Remote" denotes cross-socket accesses. DIMM latency was measured on socket 0 for both local (NUMA 0) and remote (NUMA 1) memory nodes. Similarly, CXL memory (NUMA 2) was accessed locally from socket 0 and remotely from socket 1. Additionally, we compared the latency under different access patterns, including sequential and random orders. During all measurements, the CPU hardware prefetcher was disabled to ensure accurate characterization of pure off-core memory latency.

Overall, remote memory access incurs higher latency compared to local access due to greater physical distances. Latency for CXL memory significantly exceeds that of traditional DIMM DRAM, ranging from approximately 50% to 300% higher. We observed considerable latency variation across different CXL memory devices. ASIC-based CXL memory exhibits the lowest latency, typically within the range of 200-300 ns for local access. The latency across ASIC-based vendors is similar, with ASIC-CXL-1 and ASIC-CXL-2 slightly higher than ASIC-CXL-3. FPGA CXL memory shows higher latency, reaching around the 400 ns range for local access, likely due to the lower operational frequency and suboptimal performance tuning of FPGA-based CXL controllers compared to ASIC-based solutions. The Pool-CXL and SHM-CXL devices demonstrate the highest latency, exceeding 500 ns, with a relatively smaller latency gap between local and remote access. This behavior arises because Pool-CXL and SHM-CXL are disaggregated memory architectures not directly attached to the host CPU; thus, longer physical distances contribute the vast majority of the increased latency. With the CPU hardware prefetcher disabled, CXL memory does not exhibit latency improvements for sequential access; performance remains comparable to random access scenarios.

When comparing Intel and AMD systems that use the same CXL devices, Intel SPR platforms generally achieve slightly lower latency than the AMD Zen4 architecture. This observation indicates that Intel implemented a more optimized CXL memory controller in Intel SPR, their first-generation architecture that supports the CXL protocol. However, on newer architectures, such as Intel EMR and AMD Zen5, the latency gap previously seen on ASIC-based CXL devices has disappeared. This suggests that AMD has improved its CXL memory controller design, closing the latency performance gap with Intel in the recent generation.

> **Observation 1.** The device controller implementations significantly impact CXL latency, and CPU hardware has only a marginal impact. The ASIC-based CXL memory devices have similar latency, all lower than FPGA CXL latency.

## 3.2 Bandwidth Scaling

We configured HEIMDALL to measure the bandwidth of different NUMA nodes with different numbers of CPU threads. As shown in Figure 5, the bandwidth scaling depends on the heterogeneous memories and the CPU vendor. "Local" refers to the performance when the memory device is connected to the same NUMA node as the CPU cores running the profiling, while "Remote" refers to the performance when the memory is on a different socket. For the DIMM configuration, we measured performance from socket 0 by accessing both the local NUMA node (NUMA 0) and the remote NUMA node (NUMA 1). To evaluate the CXL memory node (NUMA 2) connected to the PCIe slot of socket 0, we used cores from socket 0 for local accesses and cores from socket 2 for remote accesses.

In the case of HBM memory, the Zen4-2-HBM3 accesses HBM3 memory for both local and remote accesses, albeit on different sockets; thus, the local NUMA node (NUMA 0) and the remote NUMA node (NUMA 1) both correspond to HBM3 nodes. For the GH200-HBM3, the system consists of a CPU node and a GPU node, where the CPU accesses LPDDR5x and the GPU accesses HBM3e as their respective local memories. Consequently, the local NUMA node (NUMA 0) is associated with LPDDR5x, while the remote NUMA node (NUMA 1) corresponds to HBM3e. More details are presented in Figure 2.

### 3.2.1 Intel CPUs

In this section, we analyze both load and store bandwidth scaling results for DIMM and CXL memory across two generations of Intel CPUs and four types of CXL memory devices. Figure 5a through Figure 5o present the observed bandwidth behavior under varying thread counts. From this analysis, we identify four key observations.

First, in all systems, DIMM-based memory shows consistent and effective bandwidth scaling for both load and store operations as the number of threads increases. However, bandwidth scaling is more prominent in load operations than in store operations. For example, on the SPR-1-ASIC-CXL-1 with 4 DIMM channels, saturation is observed at 109 GiB/s (load) and 95 GiB/s (store) locally, and 61 GiB/s (load) and 60 GiB/s (store) remotely. On the SPR-2-ASIC-CXL-2, which features 8 DIMM channels, saturation reaches 208 GiB/s (load) and 175 GiB/s (store) locally, and 88 GiB/s (load) and 86 GiB/s (store) remotely. Although doubling the number of DIMM channels increases bandwidth, the gain is sublinear for both load and store paths–likely due to memory controller saturation. Notably, the increase is less pronounced for store operations (80 GiB/s local, 26 GiB/s remote) than for load operations (99 GiB/s local, 27 GiB/s remote), indicating that store paths are less scalable.

Second, even with identical memory configurations—specifically, 8 DIMM channels per socket and DDR5 at 4800 MT/s—bandwidth scaling varies depending on the CPU model. In Figures 5c and 5m and Figures 5e and 5o, the EMR-3-SHM-CXL outperforms the SPR-2-ASIC-CXL-2. For load operations, it reaches 248 GiB/s (local) and 113 GiB/s (remote), compared to 208 GiB/s and 88 GiB/s on the SPR-2-ASIC-CXL-2. For store operations, it achieves 191 GiB/s (local) and 103 GiB/s (remote), versus 175 GiB/s and 86 GiB/s, respectively. These results suggest that the EMR-3 CPU provides enhanced memory path efficiency. Additionally, UltraPath Interconnect (UPI) bandwidth on the EMR-3 system is improved by approximately 25% [15] compared to SPR-2, further contributing to its performance advantage on the remote path.

Third, bandwidth scalability varies significantly across different CXL memory devices. The FPGA-CXL exhibits poor scaling for both load and store operations. Load bandwidth saturates at 19 GiB/s (local) and 8 GiB/s (remote), while store bandwidth peaks at a single thread and degrades with more threads. This behavior is attributed to its limited configuration—only two SODIMMs. In contrast, ASIC-based CXL memory devices demonstrate more stable and higher bandwidth for both access types, maintaining saturation as thread count increases. For example, the ASIC-CXL-1 and ASIC-CXL-2, both connected to the same generation CPU, achieve similar local bandwidths, but ASIC-CXL-2 delivers approximately 6 GiB/s higher remote bandwidth in load operations, implying improved remote access efficiency.

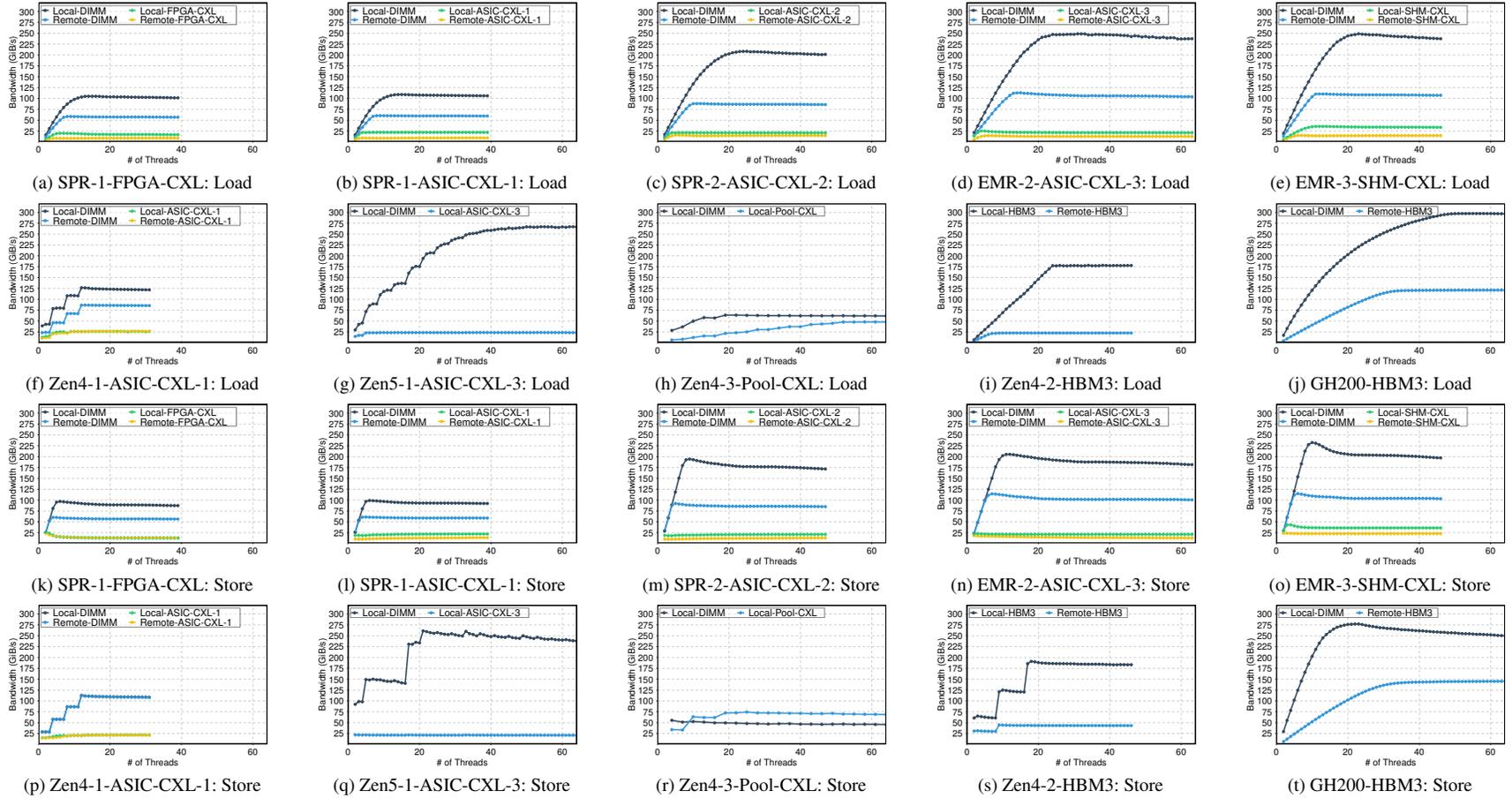

Figure 5: Bandwidth scaling when using different number of threads. The x-axis is the number of threads accessing memory, and the y-axis is the peak bandwidth achieved. We measure memory bandwidth accessing from both local and remote CPUs.

Finally, we compare systems with different CXL memory connection topologies. The EMR-2-ASIC-CXL-3, which connects CXL memory directly via the root complex, achieves 25 GiB/s (load) and 21 GiB/s (store) locally, and 13 GiB/s (load, store) remotely. Meanwhile, the EMR-3-SHM-CXL, which accesses memory through an external CXL memory pool, delivers 35 GiB/s (load), 36 GiB/s (store) locally, and 15 GiB/s(load), 23 GiB/s (store) remotely. These results exceed the theoretical maximum of a single x8 lane CXL expander (29.34 GiB/s), indicating that interleaving or efficient workload distribution within a memory pool can significantly improve total CXL memory bandwidth.

### 3.2.2 AMD and ARM CPUs

In this section, we explore both load and store bandwidth scaling results for DIMM, CXL, and HBM memory across two generations of AMD CPUs and one ARM CPU. Figure 5f through Figure 5t show the observed bandwidth behavior under varying thread counts. From this analysis, we identify three key observations.

First, DIMM memory systems demonstrate effective and consistent bandwidth scaling for both load and store operations across all AMD systems. In Figure 5f, 5g, and 5h, we observe the scaling characteristics of AMD CPUs using DDR5 DIMMs. The differences in saturated bandwidth among these systems can be primarily attributed to variations in the number of DIMM channels. The store bandwidth scaling performance is illustrated in Figure 5p, 5q, and 5r. Interestingly, in AMD systems, store bandwidth increases in a stepwise pattern. This behavior reflects the architectural layout of AMD's Core Complex (CCX), in which multiple cores within a CCX share the last-level cache, creating bandwidth jumps as thread count increases across CCXs.

Figure 5j and 5t illustrate the bandwidth scaling behavior of the GH200, which utilizes LPDDR5x memory. For load operations, it saturates at 304 GiB/s with 50 threads, showing efficient bandwidth scaling. However, for store operations, the system exhibits instability. After reaching a peak bandwidth of 283 GiB/s at 22 threads, the store bandwidth declines as thread count increases, indicating inefficiencies in the store path under high concurrency.

Second, CXL memory systems exhibit distinct bandwidth scaling behavior between ASIC-based CXL devices and the Pool-CXL. For load operations, ASIC-CXL-1 shows both local and remote CXL bandwidth saturating at approximately 26 GiB/s, with nine threads being the saturation point, as shown in Figure 5f. In contrast, Figure 5h shows that although the slope of the bandwidth increase is lower for the Pool-CXL compared to the ASIC-CXL-1 and ASIC-CXL-3, it eventually reaches a significantly higher saturation bandwidth of 49 GiB/s at 51 threads. This value is nearly twice that of the ASIC-based CXL devices, highlighting the potential of bandwidth expansion with the Pool-CXL.

For store operations, ASIC-based CXL devices demonstrate poor scaling, as shown in Figure 5p and 5q. Store bandwidth remains flat across all thread counts, showing no improvement. In contrast, Figure 5r shows that the Pool-CXL exhibits effective store bandwidth scaling: performance increases steadily with thread count and saturates at 69 GiB/s with 20 threads. This is $2.8\times$ higher than the store bandwidth observed with ASIC-based CXL devices, further demonstrating the scalability advantage of the Pool-CXL.

Finally, we explore the bandwidth scaling characteristics of HBM3 and HBM3e memory on the Zen4-2-HBM3 and GH200-HBM3. The Zen4-2-HBM3 utilizes HBM3 exclusively on both the Complex Core Dies (CCDs) and the Accelerator Core Dies (XCDs), which share memory via the Infinity Fabric interconnect to reduce redundant memory copies. We profile bandwidth scaling from the CPU side, and the results are shown in Figure 5i and Figure 5s.

For local load operations, bandwidth scaling shows efficient performance as the number of threads increases. However, the CCDs on the Zen4-2 can utilize only a fraction of the chip's total HBM3 bandwidth. As a result, the local load bandwidth saturates at 175 GiB/s with 24 threads; for reference, the full 5.3 TB/s of HBM3 bandwidth is primarily intended to serve the XCDs rather than the CPU cores. For remote load operations, although the Infinity Fabric supports a theoretical bandwidth of 112 GiB/s for cross-socket links, the observed remote load bandwidth saturates at only 25 GiB/s. The root cause of this limitation has not yet been identified.

In the store results shown in Figure 5s, the Zen4-2 exhibits stepwise bandwidth scaling similar to that of the Zen4-1 system. This behavior reflects the hierarchical structure of the CCDs: the Zen4-2 includes three CCDs, each with eight cores. Bandwidth increases occur in steps that correspond to both the number of cores per CCD and the total number of CCDs. The local store bandwidth saturates at 189 GiB/s with 17 cores, while remote store bandwidth reaches saturation at 44 GiB/s with 9 cores. As with remote load operations, remote store performance does not fully utilize the 112 GiB/s bandwidth supported by the Infinity Fabric interconnect.

The GH200-HBM3 utilizes HBM3e located on the GPU side, which appears as a remote NUMA node from the CPU's perspective. We evaluate the bandwidth scaling performance of HBM3e through the NVLink-C2C interconnect. Figure 5j and 5t present the results for remote HBM3e accesses, showing efficient bandwidth scaling. The bandwidth saturates at 125 GiB/s for load and 149 GiB/s for store operations. Although NVLink-C2C theoretically supports up to 450 GB/s of bidirectional bandwidth [16], the achieved performance falls significantly short of this peak. The underlying cause of this discrepancy remains unclear, and further analysis is required to identify the bottleneck.

> **Observation 2.** ASIC-CXL-1 shows more stable bandwidth scaling compared to FPGA-CXL. In addition, CXL memory on Zen4-1 shows better bandwidth performance than CXL memory on SPR-1, with similar saturation levels for local and remote access, while SPR-1's CXL memory saturates at different bandwidth between local and remote access.

### 3.3 Bandwidth vs. Latency

We utilize HEIMDALL to profile latency behavior as bandwidth increases across heterogeneous memory systems, including DDR5 DIMM, LPDDR5x, HBM3, HBM3e, FPGA-based CXL, ASIC-based CXL, and pooled CXL memory. This profiling allows us to characterize the latency patterns of each memory type in relation to bandwidth, and to identify the optimal bandwidth levels at which CXL memory can be used to extend system capacity without degrading overall performance. Figure 6 presents the bandwidth versus latency measurements for CXL and DIMM devices across ten types of CPU servers, comparing both local and remote memory access. The x-axis represents bandwidth, while the y-axis shows corresponding latency.

**Intel CPUs**

In this section, we explore the bandwidth versus latency behavior of Intel CPUs. We evaluate two generations of Intel processors using five types of CXL memory devices and DDR5 DIMM to observe how latency changes as bandwidth increases. Figure 6a through Figure 6e show the results for load operations, while Figure 6k to Figure 6o present the store results. As system-level bandwidth saturation was discussed in the previous section, our focus here is on the latency trends as a function of bandwidth. From our profiling, we derive two main observations.

First, DDR5 DIMM bandwidth versus latency exhibits similar patterns across all tested CPUs. The SPR-1 system, equipped with four DIMM channels, saturates at 107 GiB/s and begins to show increased latency beyond this point. In contrast, SPR-2, EMR-3, and EMR-2 systems are configured with eight DIMM channels, so they achieve higher bandwidth and show similar latency performance.

For systems with four DIMM channels, the maximum observed latency was 366 ns (local) and 654 ns (remote) for load operations, and 752 ns (local) and 635 ns (remote) for store operations. For systems with eight DIMM channels, the load bandwidth versus latency performance shows minimal variation across CPU generations, with maximum latency ranging from 254-259 ns (local) and 561-618 ns (remote). However, in store operations, the EMR-2 system exhibits a noticeable drop in bandwidth after reaching peak performance as latency increases. This effect is more pronounced compared to SPR-2 systems and can be clearly observed in Figure 6n.

Second, CXL memory expanders show diverse bandwidth versus latency characteristics depending on the device type. The FPGA-CXL device consistently exhibits higher minimum latency than ASIC-based CXL devices–$1.52\times$ (local) and $1.36\times$ (remote) higher for load, and $1.54\times$ (local) and $1.38\times$ (remote) higher for store. In contrast, the ASIC-CXL-3 shows the lowest latency for both load and store among all CXL devices, with latency $1.1\times$ (local) and $1.3\times$ (remote) lower than that of the SPR-1 system for load, and $1.1\times$ lower (local) for store. However, in terms of maximum latency, ASIC-CXL-3 shows relatively high values compared to other ASIC-based CXL devices. This is likely due to the higher core count on the EMR-2-ASIC-CXL-3, which causes greater contention. Therefore, we cannot attribute high maximum latency solely to the device's characteristics under contention.

An interesting observation is seen in the SHM-CXL system, which exhibits the highest latency across all CXL configurations, as shown in Figure 6e and 6o. Unlike other ASIC-based CXL devices that are directly attached within the server, SHM-CXL connects to an external CXL memory pool via a network card, contributing to its higher latency. Despite this increased latency, it achieves comparable maximum bandwidth and exhibits similar latency under contention. These results suggest that pooled CXL memory could be a promising solution for applications where large memory capacity and bandwidth are prioritized over minimal latency.

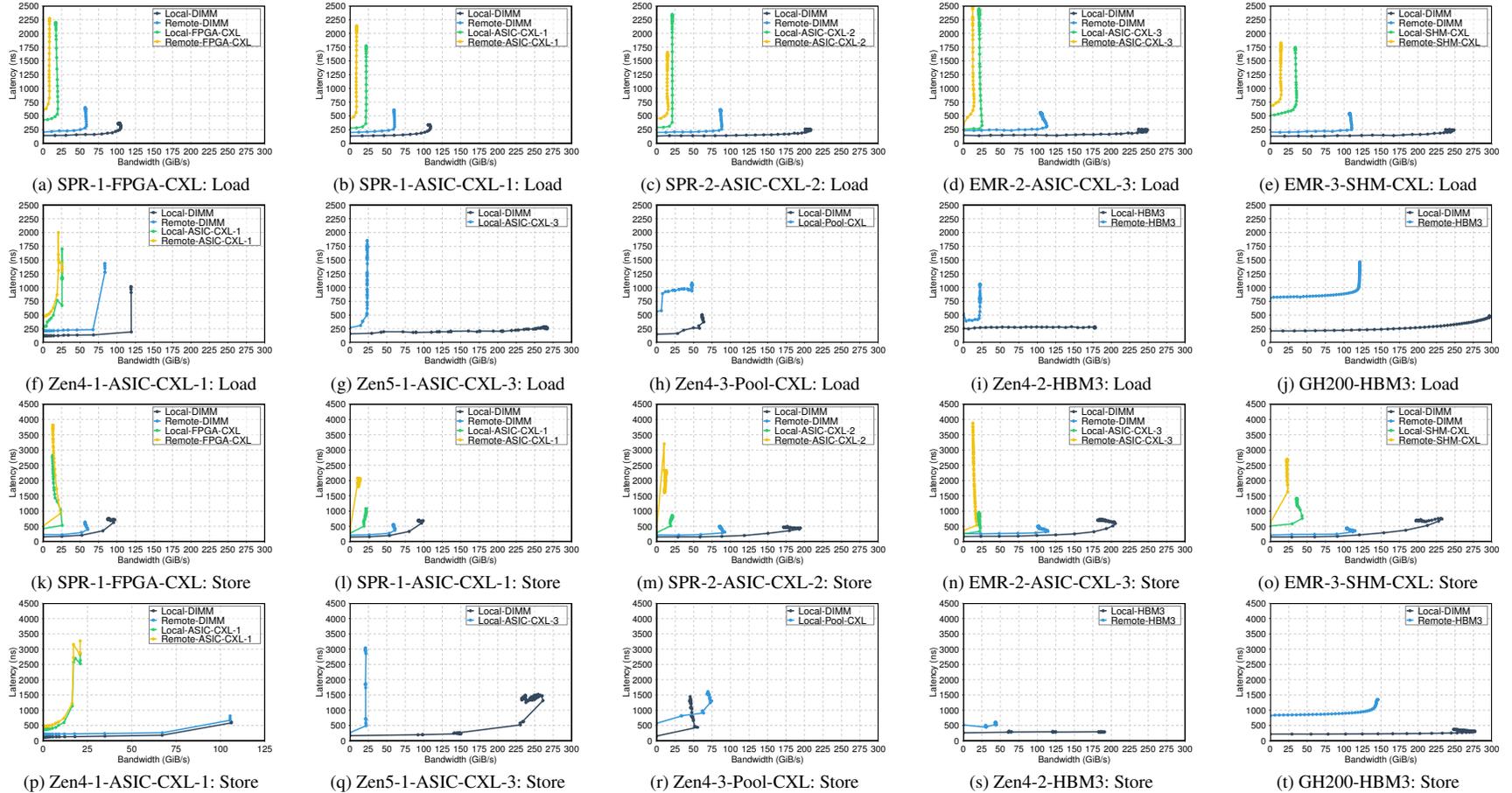

Figure 6: Bandwidth vs. latency measurements on each machine, where we measure from both local and remote CPUs, accessing either DRAM DIMM or the CXL memory.

**AMD and ARM CPUs**

In this section, we analyze the bandwidth versus latency behavior of AMD and ARM CPUs[1] using various memory types, including DDR5 DIMM, LPDDR5x, HBM3, HBM3e, and three types of CXL devices. We evaluate how latency changes with increasing bandwidth for each memory configuration.

First, DIMM memory shows varying bandwidth-latency characteristics depending on the CPU architecture. On AMD Zen4 CPUs, such as Zen4-1, the bandwidth versus latency trend remains stable for both load and store operations. As shown in Figure 6f and 6p, latency remains nearly constant as bandwidth increases. In contrast, SPR-1 exhibits a steeper latency increase in both load and store operations near peak bandwidth, as seen in Figure 6b and 6l, indicating sensitivity to bandwidth saturation.

The Zen4-2 system follows a similar trend to Zen4-1, maintaining stable latency characteristics under increasing bandwidth, as shown in Figure 6i and 6s. However, despite also using a Zen4 CPU, Zen4-3-Pool-CXL shows significantly different behavior. As illustrated in Figure 6h and 6r, both load and store latency increase sharply. We attribute this to the system using only two DIMM channels, which creates a bandwidth-constrained environment and leads to elevated latency as demand exceeds available memory throughput. The Zen5-1 processor, a newer generation AMD CPU, shows a different pattern from Zen4. As it is configured as a single-socket system, remote memory performance cannot be evaluated. In Figure 6g, the load latency remains stable as bandwidth increases, consistent with previous Zen4 results. However, the store performance differs: as bandwidth increases, latency rises noticeably, and the slope becomes steeper, as shown in Figure 6q. The GH200, based on ARM architecture and equipped with LPDDR5x memory, shows smooth and gradual latency scaling. As depicted in Figure 6j and 6t, the load latency increases gently, reaching 500 ns at 300 GiB/s. For store operations, latency remains relatively flat and only increases slightly at peak bandwidth.

Second, ASIC-based CXL memory devices exhibit generally consistent bandwidth versus latency trends across different CPU architectures. Figure 6f, 6g, 6p, and 6q present the load and store latency results for the Zen4-1-ASIC-CXL-1 and Zen5-1-ASIC-CXL-3. Despite differences in CPU generation and CXL device types, both systems show similar latency scaling behavior as bandwidth increases. They reach peak bandwidth at approximately 23-25 GiB/s, after which latency increases sharply–up to 1700-2000 ns for load operations and 3000-3300 ns for store operations. In contrast, the Zen4-3-Pool-CXL demonstrates notably different behavior compared to ASIC-based CXL memory devices. As shown in Figure 6h and 6r, the Pool-CXL maintains more stable latency scaling as bandwidth increases, and achieves both higher bandwidth and lower peak latency compared to other CXL devices. These results highlight the potential of pooled CXL memory to extend system bandwidth while maintaining lower latency than conventional ASIC-based CXL expanders.

Finally, we analyze HBM memory latency performance as bandwidth increases. Figure 6i and 6s show the results for the Zen4-2-HBM3. For local accesses, both load and store operations exhibit stable latency with nearly flat scaling. For remote accesses, however, load latency begins to rise sharply after reaching peak bandwidth, while store latency remains constant beyond the saturation point. A similar pattern is observed in the GH200-HBM3, as shown in Figure 6j and 6t. In this case, both load and store operations show a sharp increase in latency once peak bandwidth is reached.

## 3.4 Bandwidth of NUMA Interleaving

In this section, we explore methods to increase available memory bandwidth using a CXL memory expander, with a focus on maximizing bandwidth through heterogeneous memory interleaving. We first introduce basic approaches to attaching CXL memory for bandwidth enhancement, followed by methods for managing heterogeneous memory interleaving. Finally, we present profiling results for the introduced interleaving methods, highlighting the impact of varying interleaving weights between DIMM and CXL memory.

### 3.4.1 Memory Bandwidth Expansion with CXL

By integrating a CXL memory expander, we can increase both memory capacity and bandwidth, as it enables the use of CXL bandwidth in addition to traditional DIMM bandwidth. To enhance bandwidth and capacity with a CXL memory expander, two fundamental approaches are available.

---

[1] We only have one ARM CPU (from GH200) evaluated and it has HBM3, so we put it in this subsection to compare with other AMD CPUs that have HBM, our evaluated Intel CPUs do not have HBM.

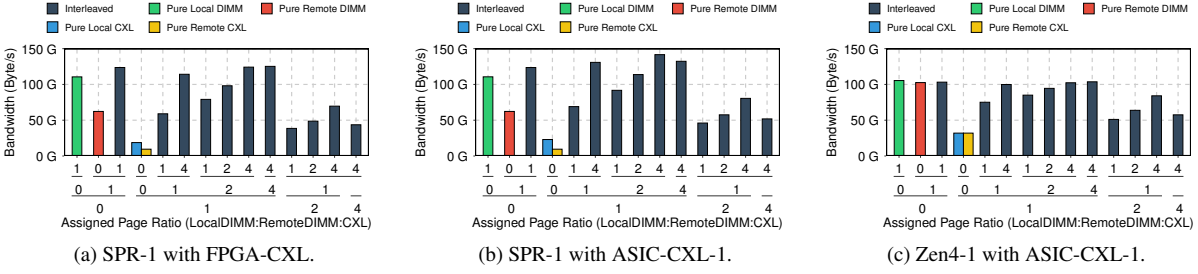

Figure 7: Bandwidth vs Weighted NUMA interleave

*CXL Direct-Attached Memory Tiering:* In this approach, the CXL memory expander is attached directly to the root complex without network devices like switches or fabrics to maximize bandwidth and capacity. This method is especially beneficial for applications constrained by bandwidth or latency. However, it is limited by the number of available CXL lanes, restricting the number of directly attached devices. This approach has been available since the CXL 1.0 specification, which supports direct attachment.

*CXL Switches and Fabric-Attached Memory Tiering:* CXL memory modules are connected through fabric and switch networks, allowing for greater bandwidth and capacity expansion but with added latency. Although this approach offers higher scalability than direct attachment, it requires a compromise due to the additional latency introduced by the fabric and switch components. This method is subject to CXL specification limitations. Single-level switch memory tiering requires devices that support the CXL 2.0 specification, while multi-level switch and fabric topology require CXL 3.0-compliant devices.

### 3.4.2 Heterogeneous Memory Interleaving with CXL

After attaching the CXL memory expander to the system using one of the methods described in Section 3.4.1, implementing efficient memory management policies becomes essential to maximize overall memory bandwidth. There are three foundational methods to interleave memory and we present our profiling results of the software-based approach.

*Hardware-Based Heterogeneous Interleaving:* This method involves configuring the system address map in the BIOS to interleave between DIMM and CXL memory expanders. However, this approach has limitations; the OS or kernel cannot control memory allocation in interleaved configurations, and applications that do not benefit from interleaving may experience inefficiencies due to the lack of control.

*Hardware + Software Based Heterogeneous Interleaving:* In this method, hardware assigns DIMM channels to different NUMA nodes, while software tools such as numactl set the interleaving ratio between DIMM and CXL memory. This approach allows the OS or kernel to manage memory allocation more effectively, but it lacks flexibility for dynamic workloads because its interleaving ratio is pre-defined instead of being dynamically adjusted at runtime.

*Software-Based Heterogeneous Interleaving:* This method enables memory allocation by configuring weights for each NUMA node at the application level. Based on these assigned weights, the kernel manages page allocation during interleaving. For example, consider a system with three NUMA nodes: NUMA0 and NUMA1 connected to DIMM, and NUMA2 connected to a CXL memory expander. The application sets weights for each NUMA node: 2, 2, and 1, respectively–in the Linux kernel at /sys/kernel/mm/mempolicy/weighted-interleave/node-number. When the application interleaves 100 pages, the kernel allocates 40 pages each to NUMA 0 and NUMA 1, and 20 pages to NUMA 2. This approach allows for flexible adjustment of interleaving weights, enabling optimized memory performance tailored to specific applications and workload characteristics.

### 3.4.3 Weighted NUMA Interleaving Results of the Heterogeneous Memory

For profiling the bandwidth of heterogeneous memory interleaving, we used the CXL Direct-Attached Memory Tiering approach in combination with Software-Based Heterogeneous Interleaving. We checked the weighted interleaving behavior on Zen4-1 and SPR-1 with ASIC-CXL-1 and FPGA-CXL based on the same system setup. Figure 7 presents the weighted interleaving result. x-axis represents the weighted interleaving ratios between local DIMM (NUMA0), remote DIMM (NUMA1), and CXL memory (NUMA2). The y-axis shows the bandwidth value.

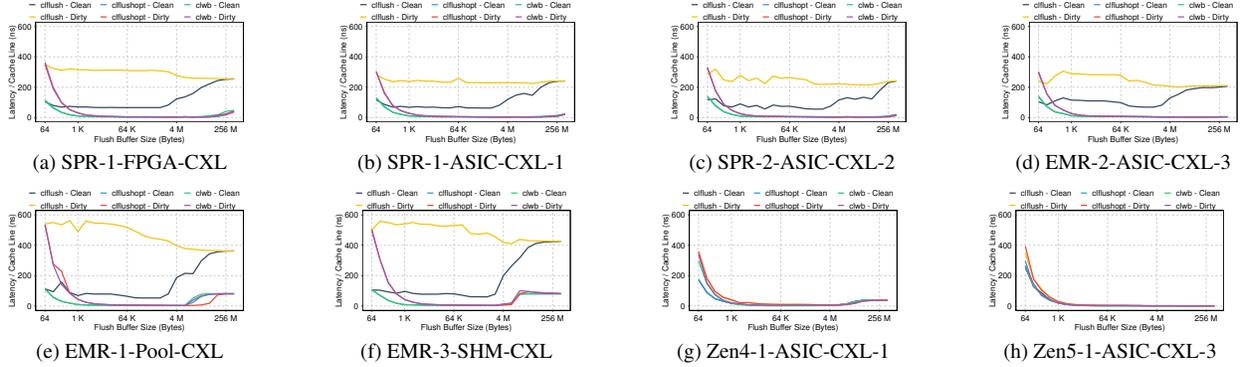

Figure 8: Flush latency when flushing different sized data, all results are from local accesses. We measured flushing either modified cachelines ("Dirty") or the unmodified cachelines ("Clean").

**FPGA-CXL vs ASIC-CXL-1.** Figure 7a and Figure 7b show the weighted interleaving results for FPGA-CXL and ASIC-CXL-1, respectively. Comparing the interleaving results reveals that ASIC-CXL-1 consistently achieves higher bandwidth than FPGA-CXL, with an average difference of 8.40 GiB/s, a maximum difference of 17.46 GiB/s, and a minimum difference of 0.04 GiB/s.

The FPGA-CXL demonstrates limited impact on memory bandwidth improvement; for instance, in the weighted interleaving configuration with DIMMs only (1:1:0), the bandwidth reaches 123.67 GiB/s, whereas the 4:4:1 configuration yields only a slight increase to 125.26 GiB/s. This suggests that the FPGA-CXL offers a minimal contribution to memory bandwidth expansion. In contrast, the ASIC-CXL-1 exhibits significantly better bandwidth enhancement capabilities than FPGA-CXL. In the DIMMs only (1:1:0) configuration, SPR-1-ASIC-CXL-1 achieves 123.63 GiB/s, increasing to 141.64 GiB/s in the 4:2:1 configuration. This suggests that ASIC-CXL-1 provides better support for weighted NUMA interleaving than FPGA-CXL.

> **Observation 3.** ASIC-CXL-1 demonstrates better bandwidth scalability with weighted NUMA interleaving compared to FPGA-CXL, achieving higher bandwidth gains across configurations.

**Zen4-1 vs SPR-1 with ASIC-CXL-1** Figure 7b and Figure 7c present the weighted interleaving results for SPR-1 and Zen4-1 with ASIC-CXL-1. The results indicate that Zen4-1 does not effectively support the weighted interleaving method. For example, when comparing the results of pure local DIMM, pure remote DIMM, and the 1:1:0 configuration, the 1:1:0 configuration shows similar bandwidth performance despite simultaneous usage of remote and local DIMMs. In contrast, SPR-1 demonstrates better bandwidth scaling when the CXL memory expander is added. The maximum bandwidth difference between SPR-1 and Zen4-1 is observed at 39.02 GiB/s in the 4:2:1 configuration. This suggests that SPR-1 achieves superior bandwidth scaling with weighted NUMA interleaving compared to Zen4-1.

> **Observation 4.** SPR-1 achieves superior bandwidth scaling with weighted NUMA interleaving compared to Zen4-1, especially when the CXL memory expander is added.

### 3.5 Flush Instructions

In this section, we evaluate CXL memory expander's performance with flush instructions by testing the latency of various flush operations (`clflush`, `clflushopt`, `clwb`) under different conditions on both Intel and AMD CPUs. The `clflush` instruction flushes all data in a cache line to memory and invalidates the cache line. Similarly, `clflushopt` flushes the cache line data to memory and invalidates it, but unlike `clflush` which is serialized, it can flush multiple cache lines simultaneously, making it more efficient for flushing large buffers (e.g., larger than several KiB). The `clwb` instruction, like `clflushopt` and `clflush`, flushes the cache, but instead of invalidating the cache line, it writes back dirty data to memory and keeps the cache line valid, thus improving performance by increasing the cache hit ratio [17].

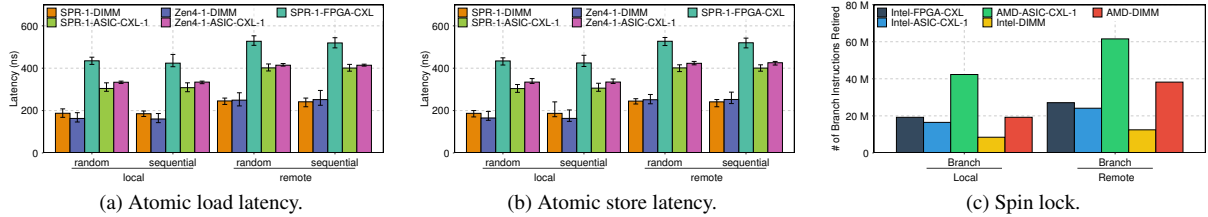

(a) Atomic load latency.   (b) Atomic store latency.   (c) Spin lock.

Figure 9: Atomic operations on CXL are slower and causes more branch instructions to be executed.

For this test, we designed a benchmark capable of measuring the latency of the three flush instructions. This framework can support both modified and unmodified cache states. The unmodified cache state refers to data that has not been changed by store operations, meaning the cache line is clean. In contrast, the modified state refers to cache lines where the data has been changed, resulting in dirty cache lines. Additionally, the benchmark allows adjustment of the flush buffer size, ranging from 64 B to 256 MB, to test performance across different buffer sizes.

Figure 8 shows the latency test results according to flush buffer size on Intel and AMD CPUs, respectively. The x-axis represents the flush buffer size in bytes, and the y-axis shows the latency per cache line in nanoseconds. These two graphs illustrate how the latency for cache line flushing changes as the flush buffer size increases.

**Intel CPUs.** Results on Intel CPUs show improved performance on both `clflushopt` and `clwb` compared to `clflush`. This is because these instructions can flush multiple cache lines in parallel, increasing the total throughput. However, flush operations exhibit varying latencies across different CXL devices, largely influenced by the read and write latencies of the respective CXL devices. Generally, flushing a clean cache line shows lower latency compared to flushing a dirty cache line. Interestingly, on Intel CPUs, we observe that as the flush buffer size increases beyond 4 MB, the latency of `clflush` on clean cache lines begins to rise, gradually approaching that of dirty cache lines.

**AMD CPUs.** The observation for AMD CPUs shows a different pattern from Intel CPUs. All the three types of flush instructions demonstrate similar latency, which indicates that `clflush` on AMD CPU can also be executed in parallel, different from the Intel CPUs. As the buffer size increases, the latency for all three flush instructions converges beyond 4 KiB.

> **Observation 5.** CXL memory supports all three flush methods. On Intel CPUs, the performance of the `clflush` instruction is inferior to that of `clflushopt` and `clwb`, as `clflush` is serialized. In contrast, AMD CPUs do not exhibit such limitations for the `clflush` instruction.

## 3.6 Atomic Instructions

In this section, we explore how the atomic instructions work with CXL system. Atomic instructions play a critical role in multi-core, multi-threaded computing environments. They guarantee that a single operation is executed without interruption, preventing other instructions from accessing the affected memory until the operation is completed. This is crucial for implementing synchronization primitives like locks, semaphores, and other concurrency control mechanisms.

With the introduction of CXL, the memory ecosystem has expanded. Since CXL.io is based on PCIe and PCIe has supported atomic operations since v3.0, we conduct experiments to verify whether atomic instructions are effectively supported on shared memory resources using CXL memory expanders. The results confirm that CXL memory has supported all the x86 atomic operations, including `CMPXCHG` and other instructions that support the `LOCK` prefix. We also measure the atomic load and store latency of CXL memory using `CMPXCHG` instructions, with all hardware prefetchers disabled during testing. The results are presented in Figure 9(a) and Figure 9(b). The latency is similar to the regular load and store latency shown in Figure 4.

When atomic operations are not natively supported, methods such as spinlock or mutex are needed to maintain synchronization. We study the potential spinlock overhead caused by CXL memory. In this test, two threads share data. Each thread writes a random value to the shared data and then executes a flush operation on it. To maintain synchronization, a spinlock is applied before the write and released after the flush. We measure the number of executed branches in the spinlock loop using the perf tool. As shown in Figure 9(c), when the data is allocated in CXL memory, there is a significantly higher branch count, indicating that the CPU spends more time spinning and waiting for the lock. This increase is due to the higher latency of CXL memory, which increases the execution time for each thread and, consequently, causes threads to spend more time in the busy-looping state while waiting for the lock. On Zen4-1, the spinlock overhead for both DIMM and CXL memory is more than 100% higher than on SPR-1, exceeding their memory latency difference, suggesting that SPR-1 may have better optimizations to reduce CPU spinning. In summary, when running workloads with frequent spinlock operations on CXL memory, CPU spinning becomes a significant issue, and strategies like increasing the sleep time within the spinlock loop can be considered to help mitigate this overhead.

Table 2: Lock-free Data Structures

| Library | Data Structure | Category |
| --- | --- | --- |
| Boost | lockfree::spsc_queue | Queue |
| Boost | lockfree::queue | Queue |
| Folly | AtomicHashMap | Map |
| Libcds | container::MichaelHashMap | Map |
| Junction | ConcurrentMap_Linear | Map |
| Junction | ConcurrentMap_Leapfrog | Map |

Table 3: Configurations of Threads and Data Location

| Configuration | Explanation |
| --- | --- |
| Same local DIMM | Both threads on the same socket. Data on the DIMM connected to this socket |
| Same local CXL | Both threads on the same socket. Data on the CXL memory connected to this socket |
| Same remote DIMM | Both threads on the same socket. Data on the DIMM connected to the other socket |
| Same remote CXL | Both threads on the same socket. Data on the CXL memory connected to the other socket |
| Diff setter CXL | Setter and getter threads on different socket. Data on the CXL memory connected to setter's socket |
| Diff getter CXL | Setter and getter threads on different socket. Data on the CXL memory connected to getter's socket |

## 3.7 Lock-free Data Structures

After discussing atomic instructions on CXL systems, we studied lock-free data structures' (LFDSs) performance on CXL systems. LFDSs are concurrent data structures that provide thread safety when multiple threads access the shared data. LFDSs use atomic instructions instead of mutual exclusion mechanisms such as mutex and semaphore. LFDSs are widely used in latency-sensitive scenarios to provide high-performance concurrent shared data access, e.g., requests and completion queue in the NVMe protocol are lock-free ring buffers. Therefore, it is critical to understand the performance of LFDSs in the CXL system and the differences compared to traditional DIMMs.

To explore LFDSs' behavior in CXL systems, we collected 6 LFDSs from 4 popular C++ libraries listed in Table 2. They cover two representative categories of data structures, which are queue for linear memory access and map for random memory access. We launch two concurrent threads accessing the data structure at the same time. For queues, one thread executes enqueue operations, and the other one executes dequeue operations, while operations are update and get for maps. Each thread performs one million operations, and we measure the time it takes for both threads to complete all. Before measuring, we initialize the data structure with different numbers of elements, so that we can look into how cache affects the performance and how the performance changes with the data structure's size. We run experiments with different thread and data placement locations listed in Table 3.

Figure 10 shows the results. Each row contains results of the same data structure, and each column represents location configuration of threads and data. In each figure, the y-axis is the total execution time the two threads spend on completing one million operations each. The x-axis is the space used by data elements, which is simply calculated by multiplying the element size by the number of elements. Queue's element size is 8 bytes and map's is 16 bytes (8-byte key and 8-byte value). In each figure, each line represents a machine with different CPU and CXL devices. Zen5-1-ASIC-CXL-3 machine only has one socket, so we can only run experiments with placing two threads on the same socket and data on either local DIMM or local CXL memory. Because benchmark of boost multi-producer multi-consumer (MPMC) queue always crashes on EMR-3-SHM-CXL machine, we cannot get the performance data from it. Some data points are missing due to out-of-memory errors.

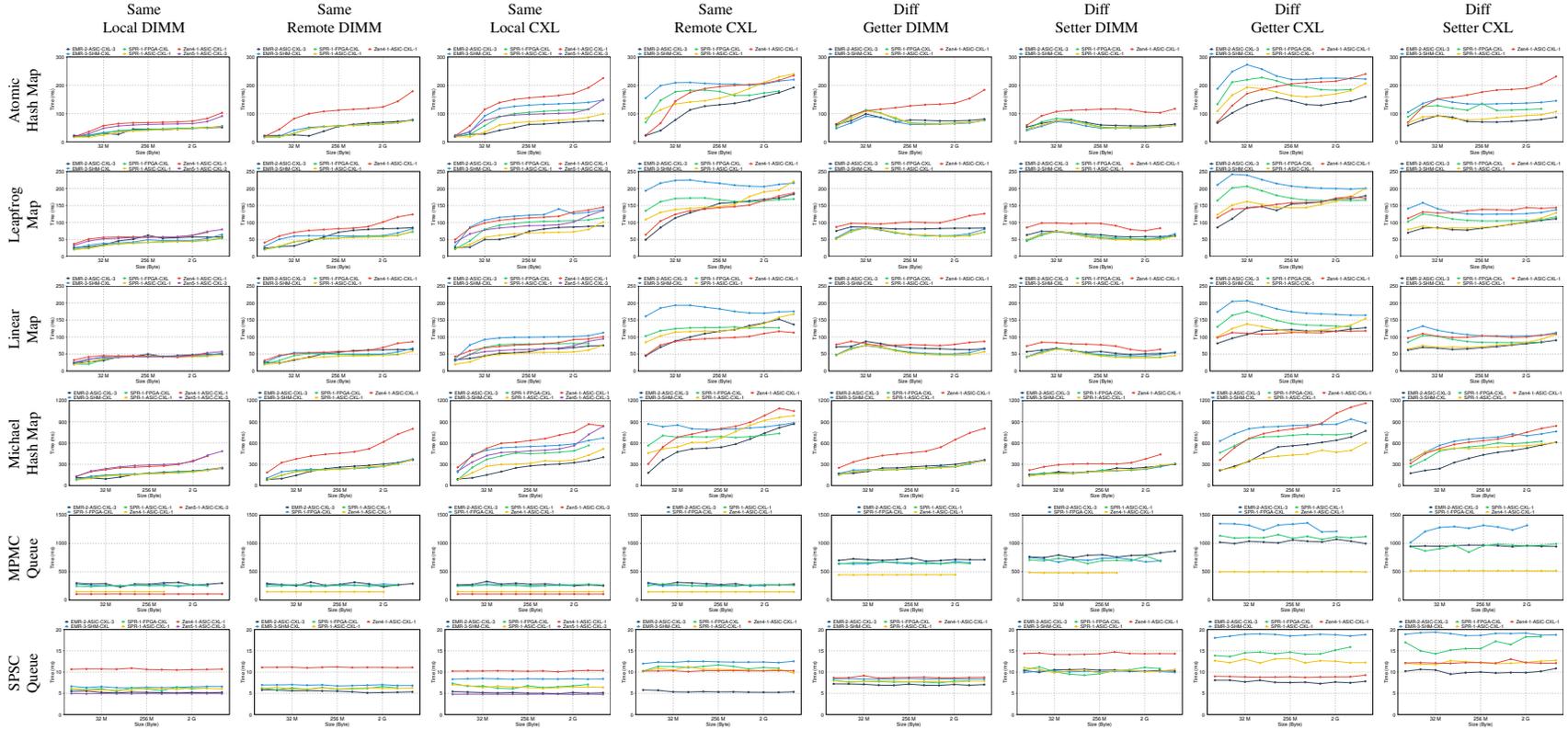

Figure 10: Lock-free data structure performance on different systems. The x-axis shows the size of all elements in the data structure, and the y-axis shows the total time to complete all operations in the microbenchmark. Each benchmark has a setter thread and a getter thread, each executing 1 million operations. We vary the thread and data location on different systems and measure performance. *Same* means both threads are on the same socket, and *Diff* means they are on separate ones. When both threads are on the same socket, the *Local* represents that DIMM/CXL memory is directly connected to the socket, and the *Remote* one is connected to the other socket. If threads are on different sockets, the *Setter/Getter* indicates that the DIMM/CXL memory is connected to the socket on which the setter/getter thread runs.

For both boost single-producer single-consumer (SPSC) queue and MPMC queue, the performance does not change with size since queue's memory access is linear. When both threads run on the same socket, on most machines, DIMM does not outperform CXL memory apparently. Cache access performance should dominate this situation. However, remote CXL memory has much more performance penalty, up to 2X, on SPSC queue on SPR-1-ASIC-CXL-1, SPR-1-FPGA-CXL and EMR-3-SHM-CXL machines. We do not observe the same change on the MPMC queue. When setter and getter threads run on different sockets, although memory access is linear, CXL memory introduces substantial overhead compared to DIMM. It shows that message communication overhead between CPU sockets is different for DIMM and CXL memory even when data is cache.

> **Observation 6.** Message communication overhead between CPU sockets is different for DIMM and CXL memory even when data is in cache.

Maps behave differently from queues. Their performance is dominated by cache or inter-socket communication when data size is small, while gradually dominated by memory access latency when data size is growing larger. However, even when data size is small and both threads run on the same socket, maps' performance on remote CXL memory with all Intel SPR CPUs is significantly worse than local CXL. At this stage, performance on local CXL memory is clearly dominated by cache access and it is getting worse as the data size increases. Maps in remote CXL memory with Intel SPR CPUs seem to not benefit from cache as much as local CXL memory. It shows that Intel SPR CPUs' cache policy is different for local and remote CXL memory. We do not observe the same behavior on the Intel EMR CPU and the AMD CPU, nor when we compare the local DIMM with the remote one. When data size is large enough that performance is dominated by memory access latency, CXL memory has considerable overhead compared to DIMM, and remote CXL memory is worse than local CXL memory. On most machines, local CXL memory clearly has higher latency than remote DIMM, but EMR-2-ASIC-CXL-3 shows that local CXL memory's performance could be close to remote DIMM. It demonstrates the possibility that the performance of the CXL memory can be comparable to that of a DIMM connected to a remote socket when the hardware is well designed.

> **Observation 7.** Intel SPR CPU's cache policy can be different for local and remote CXL memory. Local CXL memory can benefit from cache better than remote one.

## 4 Micro-Architecture Characteristics

### 4.1 Host CPU Cache Utilization of CXL Memory

Prior work has demonstrated that PCIe peripherals like NICs are only capable of utilizing a limited portion of the CPU's LLC via Direct Data I/O (DDIO) [18, 19]. Because the CXL.io protocol is built on top of PCIe, it is essential to investigate whether the same limited cache utilization also applies to CXL memory devices.

To that end, we conducted a pointer-chasing experiment designed to reveal host CPU cache utilization by CXL memory. We began by partitioning a contiguous memory region into 64-byte (one cache line) blocks. Each block is initialized with a pointer that randomly points to another block. By recursively dereferencing these pointers, all blocks are accessed in a random order. We controlled the stride size between blocks and varied the total number of blocks in the pointer-chasing task, measuring the average access latency for each block. This approach enables us to infer the underlying cache utilization and structure by observing how access time scales with both the stride size and the number of blocks. A rise in observed latency indicates an increased likelihood of cache evictions, thereby shedding light on the extent to which CXL memory can occupy and benefit from the CPU's LLC.

We conducted our experiments under both local and remote CXL access scenarios. We examined different instruction types, including regular load/store, non-temporal instructions and atomic instuctions.

**Regular Load/Store:** Figure 11 presents heat maps of read latency collected from the pointer-chasing benchmark. The deep blue regions represent cache hit latency, indicating that the corresponding stride size and block count are small enough for all data blocks to reside in the CPU cache. When moving beyond these regions, latency increases due to cache evictions. As such, the shape and extent of the deep blue areas reflect the structure and effective capacity of the cache. The first two rows in Figure 11 show the results obtained using CXL data. On both Intel and AMD CPU, when the data originates from a local CXL device, the entire LLC capacity can be fully utilized. However, for data accessed via a remote socket CXL device, all Intel machines exhibit a distinct pattern. Specifically, the maximum cache capacity usable by remote CXL data is approximately 1/8 of the LLC size on Intel SPR and 1/4 on Intel EMR. In contrast, AMD cache shows a symmetrical pattern for local and remote CXL access.

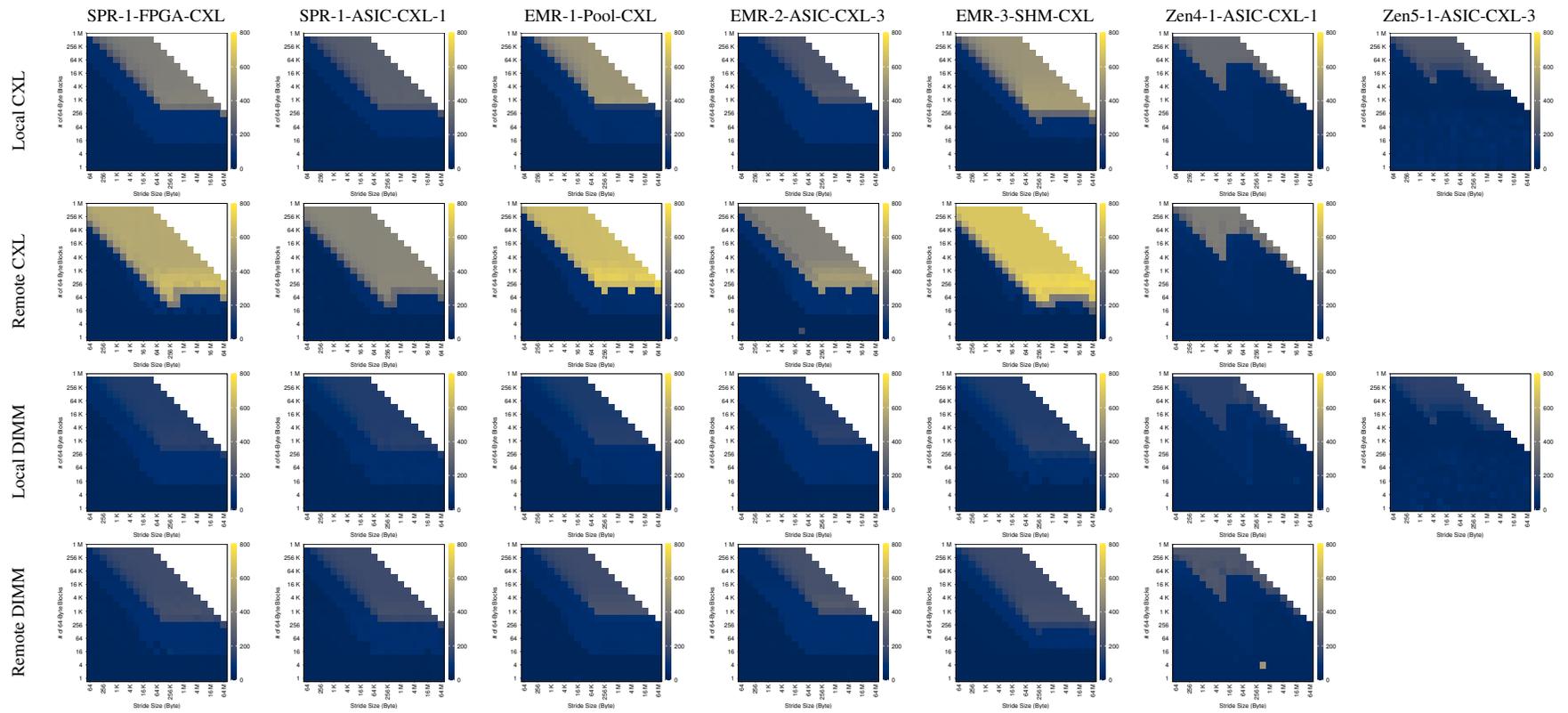

Figure 11: Cache utilization of local and remote memory read in different systems with CXL and DIMM.

We performed the same tests on DIMM as shown in the 3rd and 4th rows in Figure 11, and found both local and remote DIMM access can fully utilize the LLC on Intel and AMD CPUs. This suggests that the reduced cache capacity for remote CXL traffic on Intel machines is likely caused by the implementation of the CXL controller on the host CPU side. Furthermore, similar results were observed for regular store, and these findings remained consistent across all tested platforms. Consequently, developers should be cautious when employing remote CXL memory on Intel CPUs for data-intensive or latency-sensitive workloads, as constrained cache utilization may increase average latency.

**Non-Temporal Load/Store:** We evaluated non-temporal memory operations, which are typically employed to mitigate cache pollution for data that is not expected to exhibit high reuse. Our findings reveal that, on Intel platforms, non-temporal loads originating from either DIMM or CXL memory continue to leverage the LLC in a manner akin to regular load instructions. Consequently, the previously noted asymmetry of LLC utilization between local and remote CXL memory remains evident under non-temporal loads on Intel CPUs. In contrast, on AMD CPUs, non-temporal loads bypass the LLC entirely, relying instead on higher-level caches. For non-temporal stores, they bypass the entire cache hierarchy on all the platforms and memory devices, thereby avoiding cache utilization.

**Atomic Load/Store:** We also evaluated the cache utilization of atomic operations, and the results are consistent with those observed for regular load/store operations.

> **Observation 8.** On Intel SPR and EMR CPUs, remote CXL access is unable to fully utilize the LLC, and this constraint is independent of both the implementation of the CXL device and the types of memory operations employed.

**SNC Mode:** In SNC mode, a single CPU chip is partitioned into multiple NUMA nodes, each independently managing a region of the LLC. Our Intel CPUs support up to SNC2, which splits the chip into two sub-NUMA nodes. All the other tested CPUs have consistent results, so we use the results of SPR-1 as an example in Figure 12.

As Figure 12a shows, when the pointer-chasing task is executed on one node with local DIMM as the data source, the memory accesses are confined to that node's share of the LLC, thereby limiting utilization to half of the total LLC capacity, following the SNC partitioning.

In contrast, remote socket DIMM access and local CXL access (Figure 12b) can still use the entire LLC, indicating a distinct behavior based on the data source's locality.

In SNC (Sub-NUMA Clustering) mode, the CPU divides a single chip into multiple NUMA nodes, each managing its portion of the LLC. Our SPR-1 machine supports up to SNC2, which splits the chip into two NUMA nodes, and each node controls half of the LLC. When running the pointer-chasing task on one node, if the data comes from local DIMM DRAM, only half of the LLC is utilized, following the SNC partitioning, shown in Figure 12a. However, remote socket DIMM access (Figure 12b and Figure 12c) can still use the entire size of the LLC.

This is consistent with the previous finding [20], indicating that SNC does not enforce LLC partitioning for local CXL or remote DIMM accesses. For remote CXL, LLC usage remains limited and is shared between the two SNC nodes within the same socket. AMD implements a similar feature known as Nodes per Socket (NPS). However, in our experiments on AMD systems, we did not observe LLC partitioning effects under the NPS mode, regardless of whether DIMM DRAM or CXL memory is utilized.

**Intel CAT:** Intel CAT [21] is another mechanism available for cache partitioning. CAT allows the definition of multiple Classes of Service (CLOS), each linked to a specific bitmask. Bits set to 1 within a given bitmask designate cache ways or slices accessible to the corresponding CLOS. Cache partitioning can be achieved by assigning these CLOS masks to specific processes or CPU cores. We evaluated whether CXL memory access adheres to CAT-imposed cache partitioning. Our experiments covered various CAT masks. The result is shown in Figure 14. For example, when the LLC CAT mask is set to 0xff (the default is 0x7fff), we observe that the LLC size available to CXL memory is reduced, allowing only about half of the associativity to be used. The results showed that both local and remote CXL memory accesses consistently follow CAT-based cache partitioning rules.

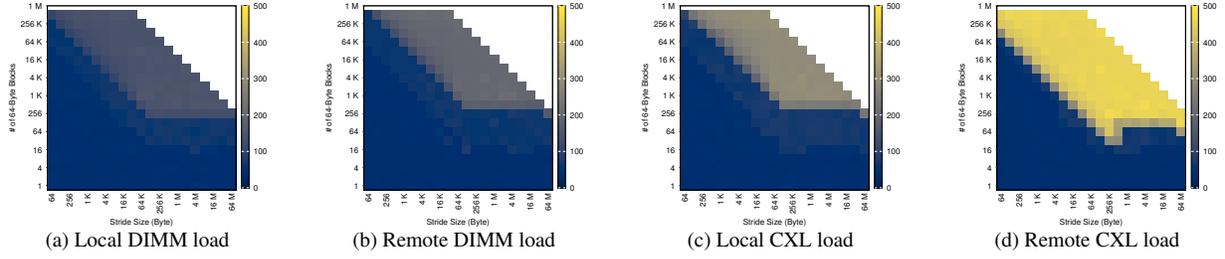

(a) Local DIMM load    (b) Remote DIMM load    (c) Local CXL load    (d) Remote CXL load

Figure 12: Cache utilization under SNC2 on SPR-1-ASIC-CXL-1

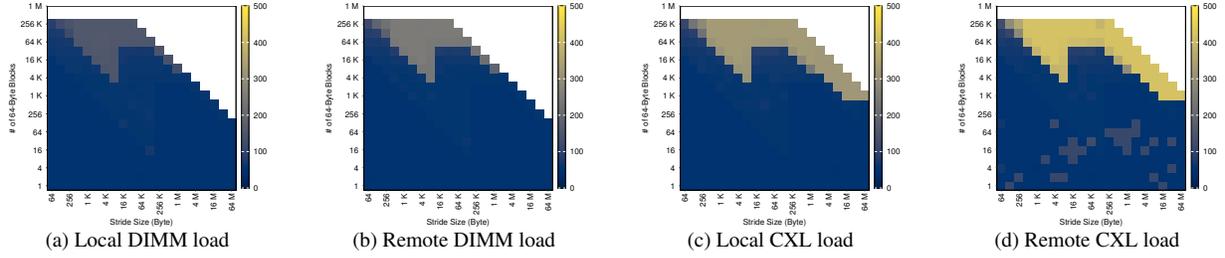

(a) Local DIMM load    (b) Remote DIMM load    (c) Local CXL load    (d) Remote CXL load

Figure 13: Cache utilization under NPS2 on Zen4-1-ASIC-CXL-1

Notably, the maximum cache utilization for remote CXL data remains limited and cannot be increased by modifying the CAT bitmask. This behavior contrasts with that of DDIO caches, where the upper limit of the capacity can be adjusted through CAT settings [19].

> **Observation 9.** Data originating from CXL memory do not follow the LLC partitioning under SNC mode. For scenarios requiring strict LLC partitioning for global data, CAT is recommended as a more effective solution.

### 4.2 Remote CXL Access Analysis

This section provides an in-depth study of remote CXL access on Intel CPUs, using (SPR-1) as an example. The analysis focuses on both the cache allocation policy and the data path analysis.

**Cache Allocation Policy:** The DDIO cache typically utilizes a fixed LLC location [18]. Given that remote CXL access also uses a limited LLC capacity, we investigate whether it follows a similar caching mechanism. First, we test the remote CXL access cache utilization with DDIO function disabled. The results still show that it has a limited LLC usable size. Then we investigate where this region is located in the LLC. In the experiment, there are two processes: P0 and P1. P0 accesses data from remote CXL, while P1 accesses data from local DIMM. Both processes run on separate physical cores, each bound a specific CAT mask.

From previous CAT tests, a mask with four consecutive '1' bits restricts the cache size to the same amount as the upper limit of remote CXL data's usable LLC size. First, we run P0 with a 4-bit '1' CAT mask and change the mask locations to see if the remote CXL LLC region can be changed by CAT. P1 also uses a 4-bit '1' CAT mask. During each test, we keep P0's mask fixed and shift P1's mask to observe where LLC eviction occurs. The results are shown in Figure 15a. We measure the average access latency of P1. For each P0's mask, the latency spike of P1 consistently appears under the exact same CAT mask with P0, suggesting that both processes share the same LLC region and cause eviction with each other. This implies that the LLC region utilized by remote CXL data can be altered by modifying the CAT mask.

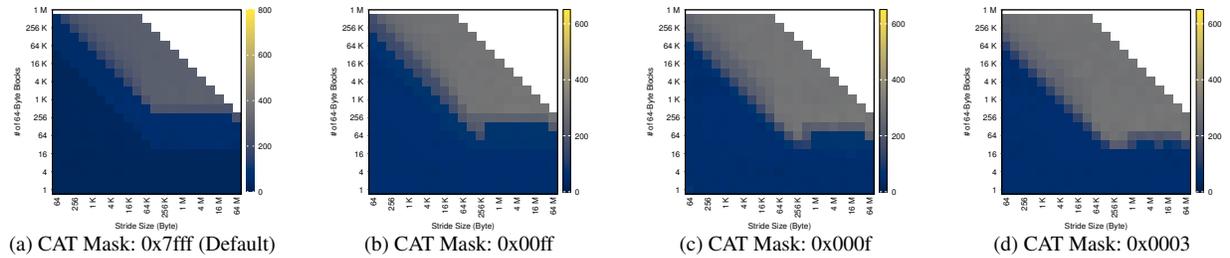

(a) CAT Mask: 0x7fff (Default)　　(b) CAT Mask: 0x00ff　　(c) CAT Mask: 0x000f　　(d) CAT Mask: 0x0003

Figure 14: Cache utilization of local CXL load under different CAT masks on SPR-1-ASIC-CXL-1

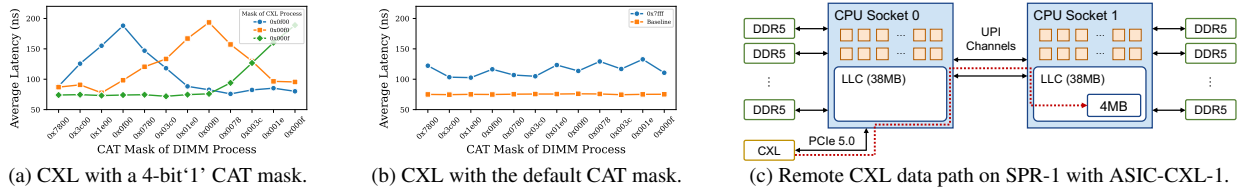

(a) CXL with a 4-bit '1' CAT mask.　　(b) CXL with the default CAT mask.　　(c) Remote CXL data path on SPR-1 with ASIC-CXL-1.

Figure 15: (a) (b) Remote CXL access with different CAT masks on SPR-1-ASIC-CXL-1. (c) CXL data path when accessing from remote CPU socket.

Next, we investigate whether remote CXL access has a preferred LLC region under the default CAT mask (0x7fff on SPR-1). The "Baseline" in Figure 15b represents the scenario where only P1 is running, resulting in all data hitting the LLC. When both P0 and P1 are running, if P0 is using a fixed cache region, we will expect a latency spike similar to Figure 15a. However, no such spike is observed during the test. While the average latency remains significantly higher than typical LLC hit latency, indicating that eviction is occurring, this suggests that remote CXL data may dynamically select its limited-size LLC region, different from DDIO cache.

> **Observation 10.** On SPR-1, the limited LLC region location for remote CXL data can be dynamically altered, different from DDIO cache policy.

**Data Path Analysis:** Next, we analyze the data path of remote CXL access. To determine whether remote CXL data traverse the Intel Ultra Path Interconnect (UPI) or only the PCIe root complex to reach the destination socket, we conduct a bandwidth test for remote CXL access while simultaneously monitoring UPI activity using Intel PCM [14]. During the bandwidth test, the UPI occupancy rate reaches 20% to 30%, indicating that remote CXL access utilizes UPI as its data path.

In addition, we analyze the cache footprint of remote CXL access. When a socket directly accesses remote CXL data, the data only stays in the destination socket's LLC rather than both sockets'. Figure 15c shows the data path of directly accessing CXL data from the remote socket.

We further explore if first placing the data in the local socket's LLC and then reading it from the remote socket could bypass the LLC size limitation for remote CXL data. First, we allocate a shared data block that is larger than remote CXL cache limit but smaller than the LLC size. Then, we run thread 0 on the local socket, randomly accessing all the shared data and recording the average access time. After multiple iterations, since all the data is stored in the cache, the average access time is below 50 ns, and the cache miss rate is low. Next, we run thread 1 on the remote socket, similarly accessing all the shared data repeatedly. The results show that when the shared data comes from DIMM DRAM (regardless of which socket it belongs to), the average access latency for thread 1 is similar to thread 0, and the cache miss rate remains low. This is because the shared data eventually gets cached in the remote socket's LLC. However, when the data comes from CXL memory, thread 1's average access latency increases to 100–150 ns, and the cache miss rate increases significantly. This is due to the remote socket being unable to store all the data in its LLC, causing some data to be evicted back to the local socket's cache.

This indicates that even when CXL data is already stored in the local socket's LLC, accessing this data from the remote socket still faces the LLC size limitation. Therefore, we can confirm that as long as the data originates from CXL memory, the LLC size limitation persists on the remote socket. We assume that this may be due to the CPU identifying the data as originating from a remote node via the physical address during the address translation process, thereby adopting a different cache mechanism.

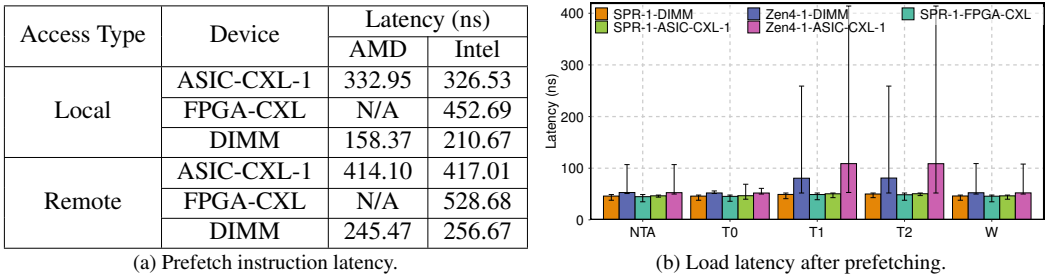

(a) Prefetch instruction latency.

(b) Load latency after prefetching.

Figure 16: Prefetching measurements: Figure 16a shows the latency of executing the `T0` prefetch instruction on different systems, and Figure 16b presents the load latency after issuing different types of prefetch instructions.

> **Observation 11.** When the remote socket directly accesses CXL data, the data are cached only in the remote socket, not both. Caching CXL data in the local socket first and then moving it to the remote socket cannot bypass the LLC size limit for remote CXL access.

### 4.3 CPU Prefetching on CXL Memory

This section presents our study of CPU prefetching on CXL memory, using both (1) software prefetching instructions which are instructions provided for programmers to hint the CPU to prefetch data, and (2) hardware prefetchers which are CPU's built-in cache prefetchers.

**Software Prefetching Instruction:** The x86 ISA supports various types of locality hints for software prefetching instructions, including T0, T1, T2, W and NTA, which fetch data to a specified location in the cache hierarchy based on the hint. According to Intel and AMD documentation, `PREFETCHT0`, `PREFETCHW`, and `PREFETCHNTA` bring data to the L1 cache or closer buffers, while `PREFETCHT1` and `PREFETCHT2` place data in the L2 cache or higher levels. `PREFETCHNTA` prefetches data to L1 cache in the LRU position or directly to line fill buffer (LFB) to avoid cache pollution, depending on the architecture implementation. This is useful for data that are accessed once and then discarded. `PREFETCHW` is a hint to the processor to prefetch data from memory into the cache in anticipation for writing.

We first measure the software prefetching latency of both CXL memory and DIMM DRAM. Figure 16a shows the latency of `PREFETCHT0`. On Zen4-1-ASIC-CXL-1, the prefetching latency for the CXL memory device is 332.95 ns for local access and 414.10 ns for remote access. For DIMM, the local access time is 158.38 ns, while the remote access time is 245.47 ns. On SPR-1-ASIC-CXL-1, the CXL memory device shows 326.53 ns for local prefetching and 417.01 ns for remote prefetching. For DIMM, the local latency is 210.67 ns, and the remote latency is 256.67 ns. Compared with the load latency in Figure 4, the performance of both CXL and DIMM software prefetching is consistent with their load latency measurements. For other prefetching hints, the latency is similar.

Next, we investigate whether the data from the CXL devices adhere to this locality mechanism. In our experiment, we first issue a prefetch instruction to fetch the data, then serialize the pipeline to ensure the prefetching completes, and finally measure the latency of a load or store on the prefetched data. The results are shown in Figure 16b. On both CXL and DIMM memory of SPR-1, we observe that the load/store latency after a `PREFETCHT1/T2` is approximately 5ns higher than other hints, aligning with the latency difference between L1 and L2 caches. This outcome matches Intel's manual [22], which states that both T1 and T2 hints bring data to the L2 cache and are implemented identically. However, on Zen4-1, we detect a discrepancy between CXL and DIMM memory. Specifically, the load/store latency after a `PREFETCHT1/T2` on CXL data is over 20ns higher than on DIMM data. We find that on Zen4-1, even when serialization is added, in some cases `PREFETCHT1/T2` does not fetch the data into the cache, and the data remains in CXL or DIMM memory, leading to a difference in overall average load latency. This phenomenon is not observed with other hints. In cases where `PREFETCHT1/T2` successfully prefetches the data, the average load latency shows 4 cycles higher than other hints on average. This result matches what is advertised in AMD's documentation [23] that the new feature of Zen4 is that `PREFETCHT1/T2` put data into L2 cache, but we currently do not have a reason why the T1/T2 are sometimes dropped.

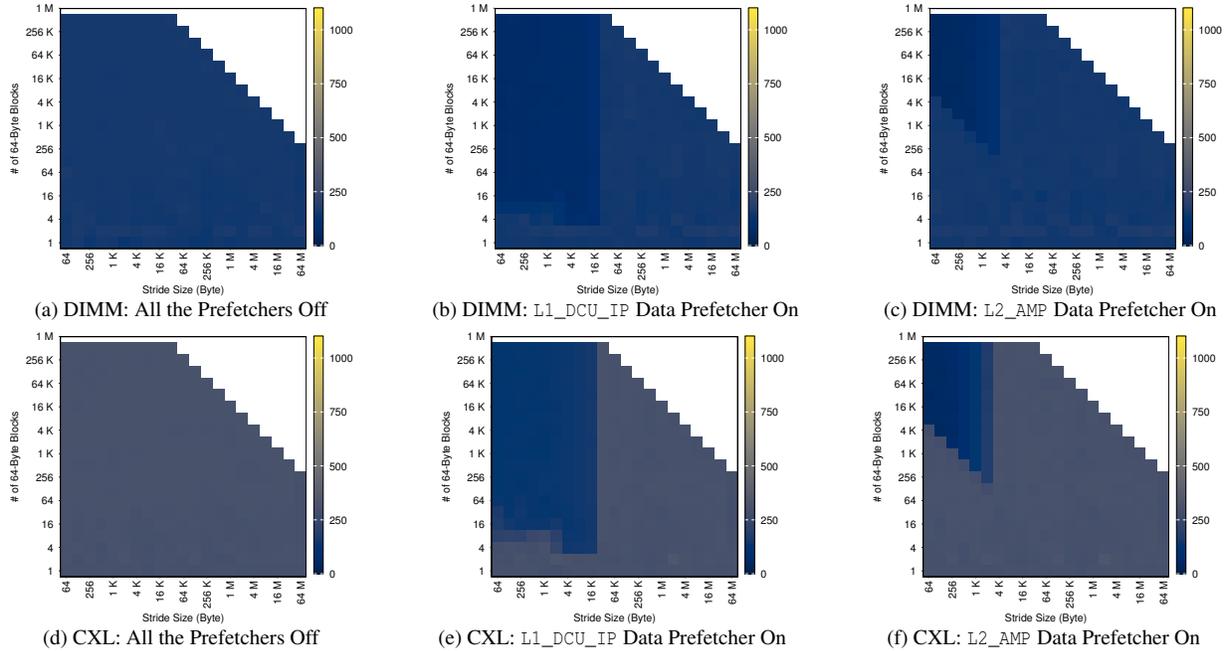

Figure 17: Hardware prefetcher footprint on SPR-1-ASIC-CXL-1

**Hardware Prefetcher:** To investigate whether the CPU hardware prefetcher performs differently during the pointer-chasing task on CXL memory, we configure the pointer-chasing access pattern to be sequential to observe the hardware prefetcher's footprint. After each access round, we also flush all test data back to memory to minimize cache noise.

Our results indicate that most CPU hardware prefetchers maintain a consistent footprint across different memory devices on SPR-1-FPGA-CXL, SPR-1-ASIC-CXL-1, and Zen4-1-ASIC-CXL-1. We illustrate two footprint examples on SPR-1-ASIC-CXL-1 in Figure 17: the `L1_DCU_IP` and `L2_AMP` data prefetchers. The `L1_DCU_IP` prefetcher begins fetching after sequentially accessing 8 64-byte blocks each when the stride is below 4 KiB. With a stride between 4 KiB and 32 KiB, it starts prefetching after 2 blocks. For strides over 32 KiB, it stops prefetching. The `L2_AMP` prefetcher, on the other hand, activates after accessing 256 KiB data sequentially but prefetches only within a 4 KiB stride range. One exception we observe is that, on SPR-1, the `L1_DCU_IP` data prefetcher only prefetches data for loads to the local socket, while on Zen4-1, the L1 data prefetcher works for both local and remote loads. These prefetchers behave the same way when using CXL as when using DIMM.

> **Observation 12.** CXL memory has supported both x86 software prefetching instructions and CPU hardware prefetchers, showing consistent results with DIMM DRAM.

## 5 Accelerator Usage

### 5.1 CXL Type 2 Accelerator Memory Performance

In this section, we study the memory access performance from the perspective of a CXL Type-2 Device. A CXL Type 2 Device is one of the most important applications of the CXL protocol, supporting CXL.io, CXL.cache, and CXL.mem protocols. This enables cache-coherent memory transactions between the CXL device and the host memory.

To support such a study, we implemented a custom CXL Type 2 device on an Intel CXL Agilex-I series FPGA with CXL support. Figure 18(a) shows an overview of the infrastructure, based on the CXL Type 2 Design Example from

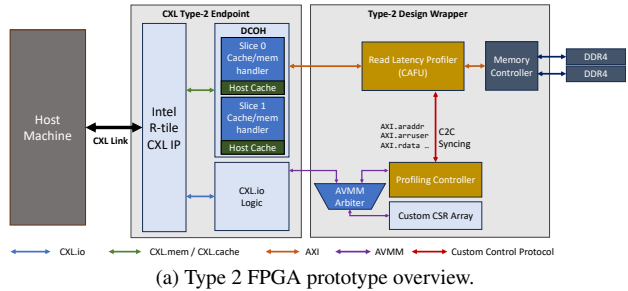 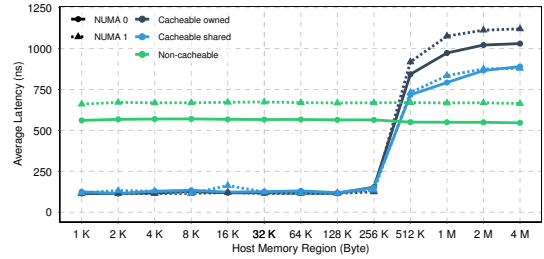

(a) Type 2 FPGA prototype overview.  (b) CXL Type 2 Device cache profiling (cache size 256KiB).

Figure 18: FPGA implementation used in studying Type 2 device memory performance

Intel. The host connects to the FPGA via the CXL Link. CXL transactions are first processed through the Intel R-Tile CXL IP, splitting all transactions into CXL.io and CXL.cache/mem protocol transactions, which are handled by the DCOH (Device Coherency Engine) and CXL.io logic, respectively.

To enable profiling, we implemented a custom latency profiler on the FPGA side as a Type 2 CAFU (Coherent Accelerator Function Unit). A CAFU interfaces with the CXL controller on the FPGA via the AXI4 protocol. The latency profiler measures the read latency of a memory request by recording the elapsed time between sending out the transaction and receiving a valid response on the AXI interface. For the host to communicate with the profiling infrastructure, we use a list of CSR (Config and Status Registers) arrays backed by FPGA registers. Every CSR register has a designated function, for example, as a command, profiling argument, and results. To initiate profiling, the host program first writes profiling arguments (host physical address, cache hint options, and etc.), and selects the read latency profiling function through the command register. This triggers the FPGA to send a host memory reading request. Upon finishing, the host program accesses the read latency through the result register.

On the FPGA, the Profile Controller module polls the content of the CSR array and performs corresponding actions. An AVMM arbiter is in place to handle the contention between the host and FPGA accessing the CSR registers. After the host program writes a read instruction in the command register, the profiling controller polls the value and triggers the Latency Profiler to start. The Latency Profiler returns a measured latency upon receiving a valid AXI response and passes the latency to the profiler controller, which subsequently writes the CSR result register. Then, the host program can read the result by memory-mapping the PCIE BAR space. Proper cross-clock domain synchronization is implemented between the Latency Profiler and Profile Controller to ensure data integrity.

### 5.1.1 Single Read Latency

Figure 18(b) shows the single read latency to the host memory initiated by the Type 2 device with no other system load. We measured across 3 request cache hints (Non-cacheable, cacheable shared, and cacheable owned). We also measure requests for local and remote NUMA nodes from the CXL device's perspective. Due to platform constraints, each request is of cacheline size (64B), and no burst read option is used. For non-cacheable reads, the latency is about 590 ns and 620 ns on local and remote NUMA nodes, respectively. For cacheable requests, cold start latency is equivalent to their non-cacheable counterparts, and then drops to about 110ns regardless of physical address. This conforms with our expectation of on-device caches as part of the DCOH. For read requests, the final cacheline state of exclusive and shared does not incur latency differences.

### 5.1.2 Device-Side Cache Size Analysis

There are two caches on the CXL endpoint of a Type 2 device: a Host Cache and Device Cache, for cache data originating from the host memory and device memory (HDM), respectively. They are not to be confused with cache on the host and cache on the device. The size of the caches is configurable by the number of DCOH (Device Coherence Agent) slices. Our setup includes 2 DCOH slices, totaling 256 KB of Host Cache and 64KB of Device Cache.

In this section, we profile the Host Cache on a Type 2 device by sequentially accessing a predefined host memory region on different NUMA nodes and using different cache hints: non-cacheable, cacheable owned, and cacheable shared. To avoid link contention, the requests are not interleaved, so a subsequent request is only issued after receiving the previous valid response. CXL.cache works on the MESI coherence protocol, allowing 4 states for cachelines: **M**odified, **E**xclusive, **S**hared, and **I**nvalid. Cacheable owned hints always invalidate the cacheline from the host cache and put the cache line in Exclusive in the device cache. The system operates in fully trusted CXL mode. During the experiment, the host provides the device with the physical address and profile region size through the PCIE interface. Then, the FPGA returns the average latency of individual accesses at the cacheline granularity. Results are shown in Figure 18(c). We observe the clear latency jump at the cache capacity of 256 KB, adhering to our Host Cache capacity with 2 DCOH slices. Cached data access times are stable around 110 ns, similar to the previous section. However, on average, read accesses using a cacheable owned hint result in about 180 ns additional overhead compared to those with a cacheable shared hint. The discrepancy is due to the extra communication required to forfeit ownership when evicting an Exclusive cacheline. Additionally, we note that cacheable requests are more expensive than non-cacheable requests, should they cause evictions.

> **Observation 13.** The cost of capacity eviction incurred by new cacheable requests of an Exclusive cacheline is much higher than that of a Shared cacheline, and both cacheable requests are more expensive than a non-cacheable read. Without a high cache hit rate, using cacheable requests as the default HDM request type for Type 2 devices is not recommended.

## 5.2 Intel DSA

Intel DSA is a hardware engine for offloading memory movement and transformation operations from the CPU cores. To understand how Intel DSA compares against `memcpy` in latency and bandwidth, we evaluate them across local DRAM, FPGA-CXL, and ASIC-CXL-1 on SPR-1. Intel DSA is configured with the parameters listed in Table 4. We use Intel DSA using Intel's Data Movement Library (DML), and all experiments were run on Node 0.

### 5.2.1 Memory Copy Latency

To understand how Intel DSA's latency compares against `memcpy`, we compare the latency of copying 512 B to 512 KiB of data using Intel DSA and memcpy. We do these experiments on a loaded and an unloaded system.

Figure 19 presents the latency comparison between Intel DSA and `memcpy` on a local memory node. When the data is evicted from the cache prior to the copy, DSA outperforms `memcpy` for transfers larger than 4 KiB. On the other hand, if the data remains cached, Intel DSA requires significantly larger transfers, around 128 KiB in our experiments, to match the performance of `memcpy`.

Conversely, for a loaded system running 10 parallel threads copying different data, we observe that Intel DSA performs similarly to `memcpy` at 16 KiB when the data isn't cached. However, when the data is cached, Intel DSA is consistently slower than `memcpy`.

For slower memories, that is, the FPGA-CXL and the ASIC-CXL-1, we observe that the break-even point between Intel DSA and `memcpy` is at 2 KiB when data is not present in the caches. If the data is already cached, `memcpy` consistently outperforms Intel DSA.

Further, when comparing copy latency for uncached data, we notice that regardless of the memory type, Intel DSA's performance over `memcpy` improves with buffer size. For example, Intel DSA is 2× faster than `memcpy` at 32 KiB, but 3.5× at 512 KiB.

Finally, when the data being copied is available in the caches, `memcpy` can directly fetch the data from the caches, making it significantly faster than Intel DSA for all memory types.

> **Observation 14.** On an unloaded system, Intel DSA can outperform `memcpy` for transfers larger than 4 KiB when using local DRAM and 2 KiB for CXL-based memories. However, if the data is already in the system caches, the break-even point can increase to as much as 128 KiB.

### 5.2.2 Memory Copy Bandwidth

Next, we will examine the total bandwidth achieved by Intel DSA and `memcpy`. For this comparison, we do not include the results for 10 threads with data eviction, as eviction introduces idle cycles for the CPU, inflating the bandwidth

Table 4: Intel DSA configuration.

| Attribute | Value | Attribute | Value |
|---|---|---|---|
| DSA Node | 0 | Number of Work Queues | 2 |
| Max Batch Size (Per WQ) | 32 | Max Transfer Size | 2 GiB |

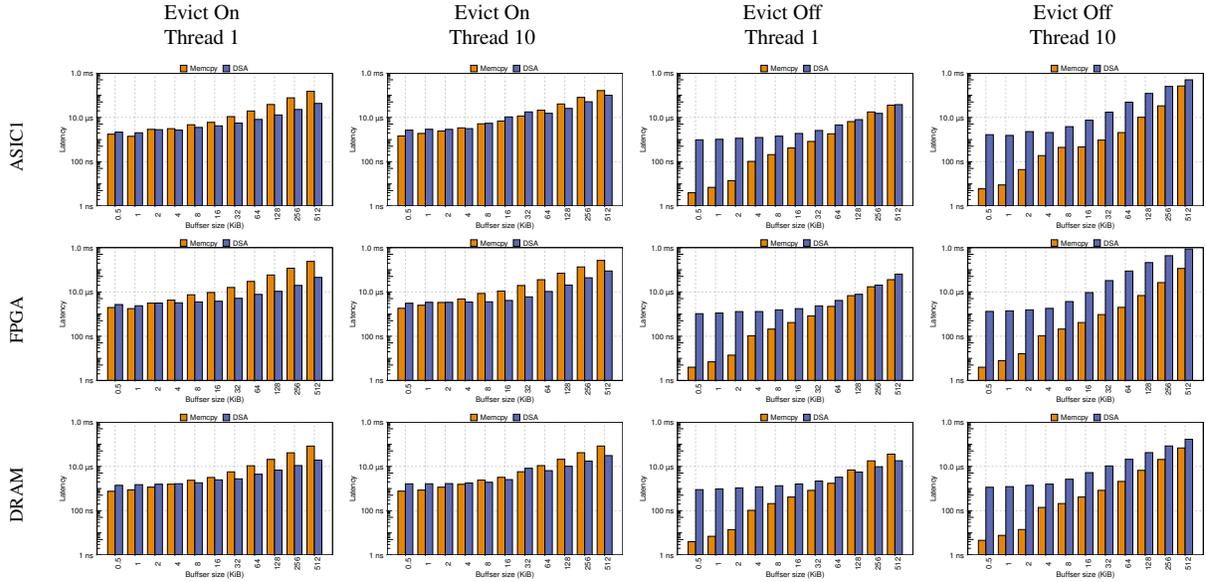

Figure 19: Latency comparison of using DSA and memcpy under different configurations.

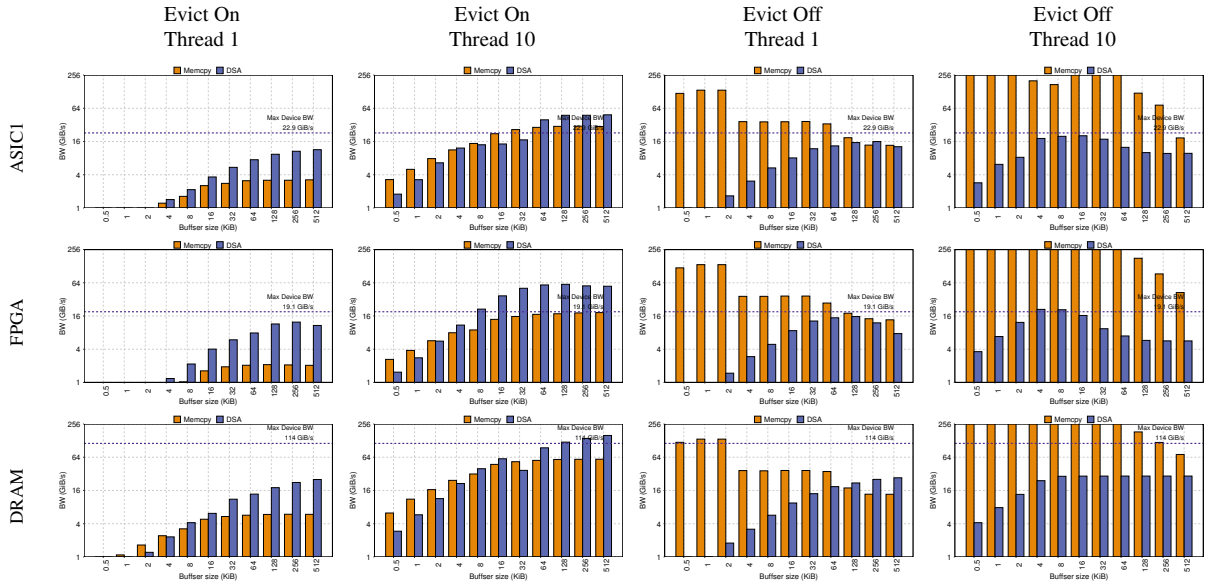

Figure 20: Bandwidth comparison of using DSA and memcpy under different configurations.

results.

Figure 20 compares the bandwidth achieved when copying data using Intel DSA and `memcpy`. With a single thread, neither Intel DSA nor `memcpy` can reach the peak device bandwidth as measured by Intel MLC [24]. However, with 10 simultaneous threads, Intel DSA reaches the peak bandwidth for slower memories (FPGA-CXL and ASIC-CXL-1), though not for the local DRAM on SPR-1. When data is cached, `memcpy` outperforms Intel DSA by a significant margin. On the other hand, when data is not cached, Intel DSA significantly outperforms `memcpy`.

> **Observation 15.** Intel DSA and `memcpy` require multiple threads to fully saturate the bandwidth of local memory and CXL-based memories. When the data is not cached, Intel DSA consistently delivers much higher bandwidth than `memcpy`, particularly for larger buffer sizes. However, when the data is cached, `memcpy` outperforms Intel DSA, although the performance difference narrows as the buffer size increases.

# 6 Applications

Based on our prior sections' profiling of CXL system architecture and accelerator performance, we further study the application performance in such systems. In this section, we study machine learning inference, graph processing, and key-value store performance on CXL systems. We observe discrepant performance characteristics when comparing CXL vs. CPU memory as well as comparing different CXL devices. We then analyze and connect such discrepancies with our architecture-level observations to explain the application-level performance characteristics.

## 6.1 Machine Learning Workloads

We study the inference performance of large language models (LLMs), one of the most popular and important workloads in today's machine learning tasks, on different CXL system configurations, using various machine learning frameworks, including PyTorch, llama.cpp, and vLLM:

- **Meta's Implementation with PyTorch [25]:** In this environment we use PyTorch with model files provided by the Meta's Llama3 repository, including `generation.py`, `model.py`, and `tokenizer.py`. We refer to this setup as "PyTorch" in the following sections.
- **llama.cpp [26]:** llama.cpp is a C++ implementation of Llama3 inference, and we evaluate the Quantized Llama3 model with it.
- **vLLM [27]:** vLLM is a high-throughput and memory-efficient LLM inference engine, and we evaluate the Llama3 inference with it.

We use the WikiText dataset [28] in these environments whenever available. For the vLLM throughput test, we use the sharedGPT v3 dataset [29] to measure performance under modern workload conditions.

### 6.1.1 PyTorch with Meta's Implementation of Llama3

We study the performance of Meta's Implementation with PyTorch and use it as a baseline to compare with the following llama.cpp and vLLM studies. In this setup, we use the testing scripts from Meta's Llama3 repo [30], including the *generator* and *tokenizer* scripts to run inferences. The input data are tokenized and processed in fixed-length segments, and the processing time is measured per each input chunk. We calculate the number of tokens generated per second to evaluate the inference performance and configure the memory devices through `numactl`. We use `numactl`'s `cpunodebind` option to bind the PyTorch process to the CPU socket assigned to a specific NUMA memory node.

**Token per second measurements.** Figure 21 shows our evaluation of Llama3 inference performance with PyTorch using different CXL machines. We first observe that inference with DRAM only is generally faster than CXL memory, and among different CXL memory devices, ASIC-CXL-1 is usually faster than FPGA-CXL, with the exception that using remote FPGA-CXL on SPR-1 is faster than remote ASIC-CXL-1. It is likely caused by the lower efficiency when using the ASIC-based CXL memory expander on the remote socket, due to the lack of native support for the CXL.mem protocol with the current generation of SPR-1 CPUs. In this case, the FPGA-based CXL.mem implementation may leverage Intel's in-house optimizations in their FPGA CXL IP to provide better remote access performance.

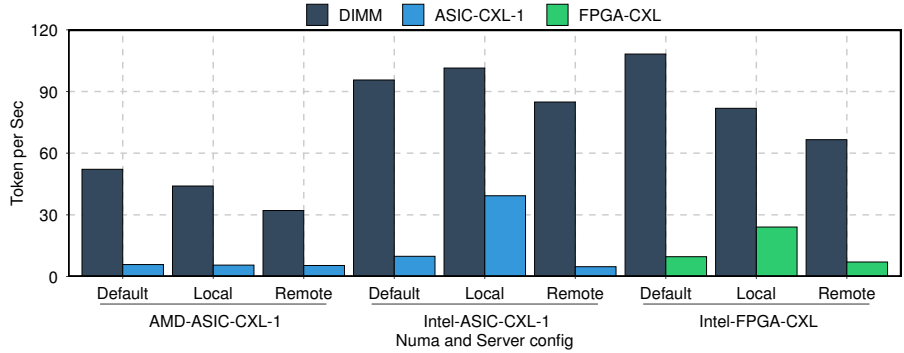

Figure 21: Token per second with PyTorch on different CXL machines. The "Default" bars show the performance with NUMA memory bind to the specific memory device, while without CPU bind (i.e., CPUs are free allocated). The "Local" and "Remote" ones have CPUs binding to corresponding CPU sockets.

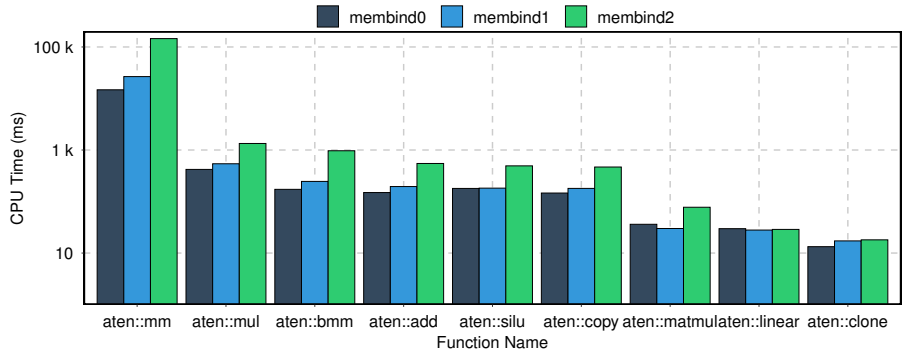

Figure 22: PyTorch profiler's results on SPR-1 with FPGA-CXL.

When using different CPU binding settings, we find that with SPR-1 and ASIC-CXL-1, binding CPUs (*Local* and *Remote*) offers higher performance using DRAM, than allowing CPUs to be freely allocated on both sockets (*Default*). As shown in Figure 21, binding SPR-1 CPU for local CXL access performs significantly better than other settings. This indicates that binding CPUs through numactl helps PyTorch better schedule tasks and allocate memory when using slower devices, such as CXL memory. In contrast, we do not observe similar performance characteristic on Zen4-1, whose overall performance is lower than SPR-1, potentially because PyTorch under-utilizes the CPU and memory performance on Zen4-1.

**Function-level profiling.** We use the PyTorch profiler to get high-level performance characteristics of the Llama3 inference. The tool reports the execution time of each internal functionality. As shown in Figure 22, using CXL memory incurs significantly longer execution time for memory-intensive operations, especially in matrix multiplication functions including *aten::mm*, *aten::matmul*, and *aten::mul*. These operations access the matrix data stored in off-core memory and incur longer execution time if CXL memory is used instead of DRAM.

> **Observation 16.** CXL can cause higher cache miss rates even with the same model inference on the same framework.

### 6.1.2 llama.cpp

In this section, we evaluate the performance of the language model using a perplexity benchmark in the llama.cpp environment. Specifically, the model used is Meta-Llama-3-8B.Q4_K_M.gguf, with 8.03 billion parameters in GGUF

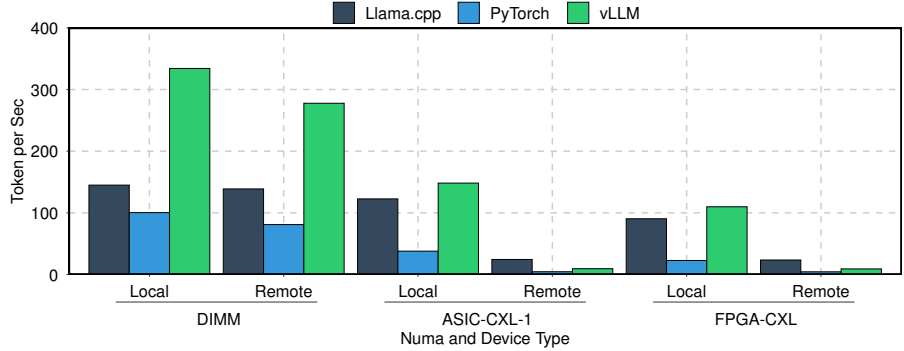

Figure 23: Token per second (TPS) comparison between all three frameworks on various machines with different memory types.

V3 format. The perplexity benchmark is run using the WikiText dataset, with a context length of 512 tokens, a batch size of 2,048 tokens, and a sequence count of 4. We also enable SMT threads and SIMD optimizations to fully utilize the hardware resources.

Figure 23 shows the tokens-per-second performance under different memory and CPU binding configurations on both SPR-1 and Zen4-1 CPUs. If *local* or *remote* is specified, we bind the model to one CPU socket and allocate all the data in its *local* or *remote* memory devices. Otherwise, no binding is enabled, and all sockets and memory devices are utilized.

Compared to PyTorch Inference of LLaMA, Llama.cpp shows less performance degradation when using CXL memory. When binding the model to one socket, SPR-1 performs better on both DIMM and CXL. However, when running without socket binding–allowing the model to utilize both sockets–Zen4-1 outperforms SPR-1 on both DIMM and CXL. Specifically, Zen4-1 achieves up to a 191.97% increase in performance on DIMM and a 241.11% increase on CXL. In contrast, SPR-1 scales less efficiently with more computing resources, with a performance increase of only 66.76% on DIMM and 75.50% on CXL. This suggests that Zen4-1 benefits from better socket interconnect performance and a more efficient scheduling policy, consistent with the findings of Section 3.2.

For local and remote access comparison, on SPR-1, the performance difference on DIMM memory is slight (Local 51.76 vs. Remote 50.48, a 2.47% decrease). In contrast, on Zen4-1, there is a 19.59% decrease on the remote DIMM. However, when using remote CXL, SPR-1 shows a much larger performance drop (25.55%) compared to Zen4-1 (19.98%). This larger drop on SPR-1 could be due to lower bandwidth and a smaller usable LLC size for remote CXL accesses, as shown in previous sections.

When comparing socket binding and no binding, unlike PyTorch inference, using CXL without socket binding even outperforms using local DIMM or remote DIMM with binding. This suggests that the Llama.cpp workload is more sensitive to CPU resource availability than to memory latency.

### 6.1.3 vLLM

In this section, we evaluate another inference framework, vLLM: a library and platform designed for efficient inference and serving of LLMs. vLLM is widely recognized for its memory-efficient operations, which guarantees high performance by optimizing memory utilization and batching requests. We deploy the same Llama3 8B model and run the benchmark_throughput.py script provided by the library on the SPR-1 with ASIC-CXL-1. The dataset is ShareGPT_v3. The key value (KV) cache size is set to 30 GiB. We use the same measurement as in previous sections and compare different memory and CPU configurations.

The results are presented in Figure 23. We observe a much higher throughput on vLLM than on Pytorch and llama.cpp on DIMM and local CXL. Local CXL shows a decrease of around 60% compared to local DIMM, which is similar to their difference in latency. However, remote CXL shows significantly lower performance, with a throughput of only 9.81 tokens per second. This indicates that besides latency, cache size and bandwidth limits could also impact remote

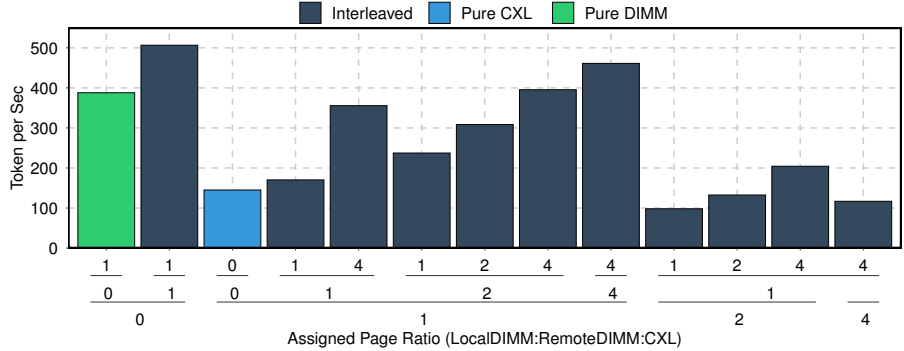

Figure 24: vLLM token per second with different weighted interleaving settings on SPR-1-ASIC-CXL-1. The X-axis shows the page assignment ratio, which, from top to bottom, represents local DIMM, remote DIMM, and local CXL. We interleave the memory while allowing CPU processes to be freely allocated across both CPU sockets.

CXL performance. This also harms performance under the default strategy (without CPU binding), because some threads access the CXL memory from the remote socket. This shows that vLLM is sensitive not only to memory latency but also to bandwidth and CPU cache size.

These results suggest that when running vLLM on CXL memory on the SPR-1, applying CPU binding to the local socket can mitigate performance drops caused by remote CXL access. In contrast, for DIMM configurations, maximizing CPU utilization without CPU binding can achieve better performance.

#### 6.1.4 Weighted interleave

We study the vLLM performance on the SPR-1-ASIC-CXL-1 by enabling the weighted interleave through the numactl option. The weighted interleave feature allows memory to be assigned, stored, and accessed across various NUMA nodes in proportions specified by assigned weights. In contrast, traditional interleaving distributes memory access evenly in a round-robin fashion across nodes. We have studied the interleaving bandwidth with microbenchmarks in Section 3.4 and this section focus on vLLM inference when using weighted interleaving.

As shown in Figure 24, we find that using pure DRAM, especially when using DRAM DIMM from both sockets, achieves higher tokens per second than using pure CXL memory, which aligns with our results in Section 3.4. In addition, interleaving both DIMM and CXL achieves higher performance than using only the CXL, where the 4:4:1 config (four pages on local DIMM, four on remote DIMM, and one on local CXL) achieves nearly the peak DIMM bandwidth. Since this system only has one memory expander, we expect the optimal configuration to change when more expanders are added.

We also find that the weighted interleaving may not fully utilize the system's aggregated bandwidth, where the added CXL bandwidth should help achieve higher than pure DIMM bandwidth. This is potentially due to the machine learning inference framework not specifically optimized to utilize the added CXL bandwidth, and also the interleaving scheme does not consider the memory access pattern. While achieving sub-optimal performance, the weighted interleaving is a drop-in solution, i.e., not requiring any software change, to expand the total memory capacity. We envision that weighted interleaving could serve as an early exploration of software that uses CXL memory and later guide the optimization strategies to fully utilize the total bandwidth.

#### 6.1.5 Hybrid Memory Placement for GPU-based Inference

We conducted our experiments on the SPR-3-ASIC-CXL-1 using vLLM with LLaMA 3 70B model. We utilized the same dataset as in previous experiments. The LLaMA 3 70B model has a total weight size of approximately 130 GiB, which exceeds the VRAM capacity of a single H100 GPU (80GiB), requiring weight offloading support. vLLM provides CPU offloading functionality that allows storing portions of the model weights in CPU memory. During inference, when calculations require weights from offloaded layers, vLLM temporarily copies these weights to reserved GPU memory space, performs computations, and then frees the space–all computations still occur on the GPU.

Table 5: Comparison of inference throughput with various offload configurations across H100 (DIMM/CXL) and GH200

| Offload Size | Token per Second | | |
|---|---|---|---|
| | H100 DIMM | H100 CXL | GH200 |
| 70 GiB | 66.72 | 25.0 | 795.03 |
| 80 GiB | 121.43 | 45.65 | 772.65 |
| 90 GiB | 131.42 | 49.18 | 711.13 |
| 100 GiB | 117.0 | 43.68 | N/A |

Table 6: CPU-to-GPU data transfer performance: Comparing DIMM and CXL memory using pinned CUDA memory

| Transfer Size | Duration (ms) | | Throughput (GiB/s) | |
|---|---|---|---|---|
| | DIMM | CXL | DIMM | CXL |
| 160 MiB | 3.01 | 8.47 | 51.72 | 18.44 |
| 128 MiB | 2.41 | 6.78 | 51.75 | 18.38 |
| 896 MiB | 16.86 | 47.28 | 51.63 | 18.38 |
| 448 MiB | 8.44 | 23.59 | 51.63 | 18.38 |
| **Average Throughput** | | | **51.68** | **18.39** |
| **Performance Ratio (DIMM/CXL): 2.81×** | | | | |

The KV cache size plays a critical role in LLM inference performance as it determines how many concurrent requests can be processed. When using vLLM, the system allocates memory for model weights and KV cache based on available GPU VRAM, typically reserving about 90% of total GPU memory for stability. For an H100 GPU with 80 GiB VRAM, this results in approximately 72 GiB of usable memory. For example, when offloading 70 GiB of weights to CPU memory, around 60 GiB of weights must remain in GPU VRAM (since the total model size is 130 GiB). This limits the available memory for KV cache to about 10 GiB, restricting the number of inference requests that can be processed. When deciding how much to offload, there is a tradeoff: offloading more weights provides more space for KV cache but increases the frequency of memory transfers between CPU and GPU.

As shown in Table 5, the H100_DIMM case achieves approximately 2.7 times higher performance than the H100_CXL case. With a significant portion of model weights stored in CPU memory, computations require frequent data transfers to the GPU. The performance gap can thus be understood through the different data transfer capabilities of DIMM and CXL memory systems. To better understand this relationship, we used NVIDIA Nsight Systems to capture GPU traces, as detailed in Table 6. This analysis shows memory copy operations occurring between kernel executions during inference. The data indicates memory copy from DIMM to GPU achieves an average throughput of 51.68 GiB/s, while memory copy from CXL to GPU is limited to 18.39 GiB/s, a 2.81x difference. This bandwidth difference helps explain the 2.7x inference throughput difference observed between the H100_DIMM and H100_CXL cases.

Our analysis indicates that during inference, memory copy operations consumed more than 99% of the total execution time. We can conclude that memory bandwidth between CPU and GPU represents the primary bottleneck for LLM inference when model weights are offloaded. The similar ratios between memory copy performance (2.81x) and inference throughput (2.7x) suggest a near-linear relationship between memory bandwidth and LLM inference performance in weight-offloaded scenarios.

Our measurements show a performance pattern across all configurations where throughput reaches a maximum at a specific offload size and then decreases. For the H100 configurations, optimal performance is reached at 90 GiB offload, while the GH200 shows its best performance around 70 GiB. This pattern can be attributed to two competing factors. When the offload size is too small (less than 70 GiB), there is insufficient memory allocated to the KV cache, limiting the number of concurrent requests that can be processed and thus reducing throughput. Conversely, when the offload size exceeds the optimal point (above 90 GiB for H100), the GPU spends more time waiting to fetch model weights from CPU memory. In this scenario, while more space is available for the KV cache, its utilization decreases due to frequent memory copy operations, consequently reducing throughput. These observations suggest that when designing LLM inference systems, one should carefully plan model weight offloading strategies by considering the hardware's memory capacity and memory bandwidth.

To further investigate the impact of memory bandwidth limitations, we conducted the same experiments using the NVIDIA GH200 Grace Hopper Superchip [16], which integrates Hopper GPU architecture with Grace CPU. This advanced superchip features 96GB of HBM3e memory and NVIDIA's NVLink Chip-2-Chip interconnect, facilitating high-speed communication between the CPU and GPU. The NVLink interconnect provides approximately 900 GB/s of memory bandwidth (with 419 GB/s from CPU to GPU and 371 GB/s from GPU to CPU), which is about 8 times faster than the PCIe Gen5 interface that delivers around 110 GB/s.

As shown in Table 5, this high memory bandwidth results in dramatically improved inference performance, with the GH200 achieving nearly 16 times higher throughput than H100_CXL and 6 times higher than H100_DIMM at comparable offload sizes. This substantial performance gain emphasizes the critical role of memory bandwidth in LLM inference with offloaded weights.

## 6.2 Vector DB

Vector databases are built for managing and querying high-dimensional data, which is essential in applications involving machine learning and AI. These systems store vector embeddings-mathematical representations of data such as text, images, and audio generated by machine learning models. By leveraging vector similarity search, they enable efficient retrieval of data which is important for tasks such as recommendation systems, image similarity searches, and natural language processing. We use two vector databases, Qdrant [31] and Milvus [32], to evaluate the CXL performance impact relative to DIMM performance. Our goal is to understand how CXL handles high-throughput, real-time similarity search workloads and the impact on Requests Per Second (RPS), precision, and latency (p95 and p99) relative to DIMM. We use *glove100* [33] as the input for both vector databases, which includes pre-trained global vectors for word representation (GloVe) for approximate neighbor search.

Figure 25 and Figure 26 shows the results from SPR-1-FPGA-CXL and SPR-1-ASIC-CXL-1 machines. For Qudrant, DIMM achieved up to 3.5× higher RPS than CXL, with the Intel-CXL machine showing the smallest RPS difference between DIMM and CXL. For the two other machines, we see a similar trend in the CXL performance. Regarding latency, the p95 and p99 latencies are below 0.1ms for DIMM across all systems, while the CXL latencies in the Intel-CXL machine show the best performance. On the other hand, the AMD-CXL and Intel-FPGACXL machines show similar CXL latencies, with an up to 8× increase for both latencies.

For Milvus, across all these machines, we observe that DIMM and CXL have similar performance in terms of RPS, with the CXL performance dropping faster as the precision increases. In addition, the Intel-CXL machine achieves the smallest max RPS for both DIMM and CXL. Regarding latency, all systems display similar p95 and p99 latencies, with the Intel-CXL machine having the lowest latency for DIMM, and Intel-FPGACXL displaying the lowest p99 latency for CXL.

In conclusion, we notice that the performance of each vector database varies and depends heavily on the underlying system, including the CPU and CXL architecture. For Qdrant, the Intel-CXL and AMD-CXL machines offer better performance, while Milvus achieves the best performance for the Intel-FPGACXL machine.

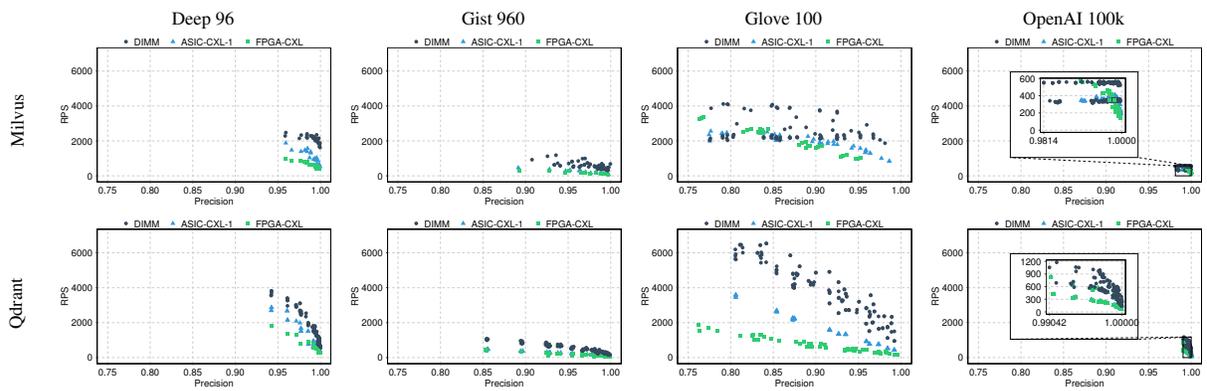

Figure 25: Requests per second (RPS) vs. precision with different memory devices.

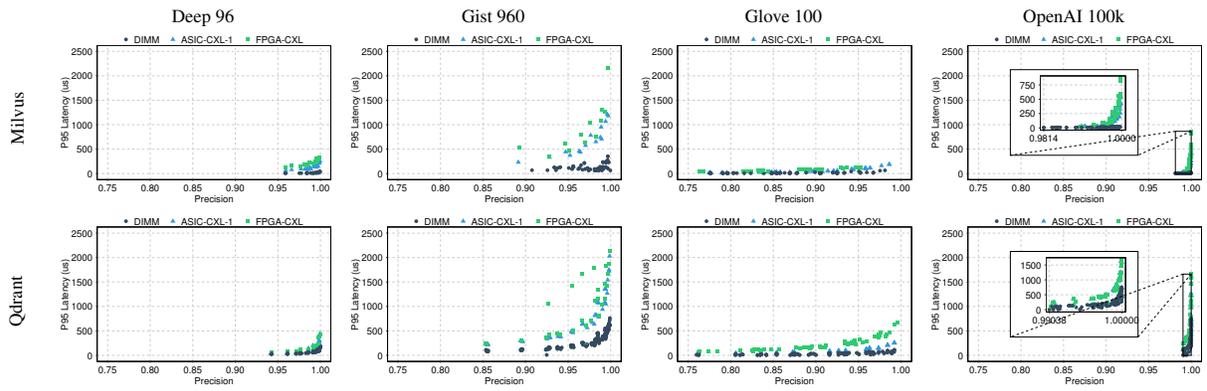

Figure 26: P95 latency vs. precision with different memory devices.

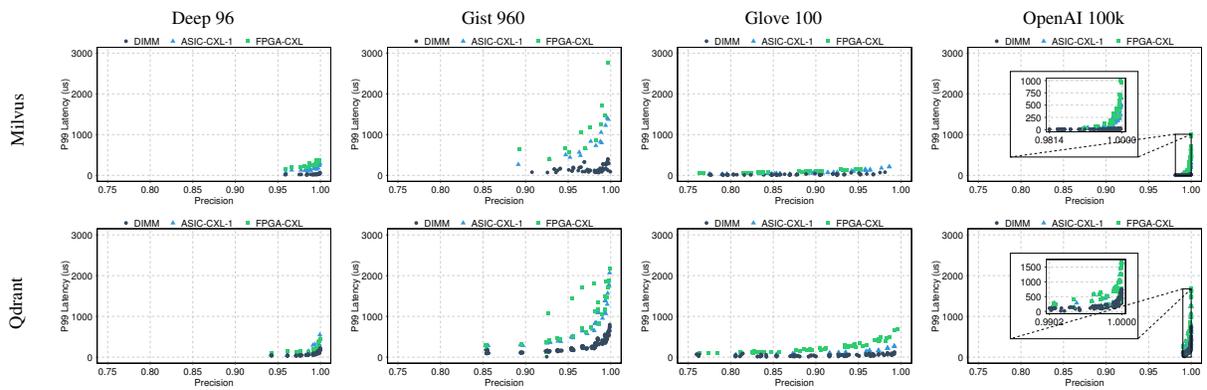

Figure 27: P99 latency vs. precision with different memory devices.

## 6.3 Graph and Key-Value Workloads

This section contains the performance of four graph (*graph generation, pagerank, local clustering, BFS*) and four KV workloads (*redis SET, redis GET, memcached SET, memcached GET*). Table 7 contains the graph and KV workloads that we used for our evaluation on Intel's and AMD's DRAM and CXL. We run these benchmarks on bare metal and a virtual machine, while we also present the results for using base pages (4 KiB) and huge pages (2 MiB) on bare metal. For all experiments we measure the execution time 5 times and report the normalized average execution time.

We use QEMU/KVM [34] version 4.2.1 to spawn and manage virtual machines (VMs) in our experimental setup. For the graph workloads we use the graph-tool module in Python and for the KV workloads we are using the redis (redis-py) and memcached (python-memcached) python clients [35, 36].

### 6.3.1 Bare Metal and Virtual Machine performance

Figures 28a, 29a and 30a show the performance for the 8 workloads that we run on bare metal and on a virtual machine, plus the geomean of all workloads. The performance is normalized against the baseline, which is each server's DIMM bare metal performance. We have the following observations: **First**, the VM overhead over the bare metal execution is proportional for both DIMM and CXL in most workloads, with the AMD-CXL machine having the lowest VM performance degradation. **Second**, the graph workloads have a worse performance overhead with CXL than the KV workloads. This is expected as the former are more memory intensive, with random memory accesses all over the address space, with *pagerank* and *local_clustering* incurring the highest performance overhead. **Third**, the best CXL performance is achieved by Intel-CXL with AMD-CXL and Intel-FPGACXL showing similar performance.

### 6.3.2 Bare Metal and Virtual Machine performance

Figures 28b, 29b and 30b presents the performance of 8 different workloads, along with the geometric mean, using both base pages (4 KiB) and huge pages (2 MiB). The performance is normalized against the baseline, which is each server's DIMM bare metal with 4 KiB pages performance.

We have the following observations: **First**, we notice that the use of 2MiB pages has a different performance impact in each machine. The Intel-CXL's geomean shows a similar performance between the two different page sizes, while the AMD-CXL's geomean shows a 60% performance degradation when using CXL with huge pages. Last, the Intel-FPGACXL's geomean shows similar performance improvement in DIMM and DRAM with 2 MiB pages. **Second**, we notice that again the worst performance degradation appears in graph workloads, due to their memory intensiveness. **Third**, the CXL performance degradation with 2 MiB pages in some cases is bigger proportionally compared to the DIMM's 2 MiB overhead (e.g., *pagerank* in AMD-CXL).

Table 7: Graph and KV workloads

| Workload | Description |
| --- | --- |
| graph_gen | Graph generation with 50M vertices and 250M edges |
| pagerank | Pagerank algorithm on the generated graph |
| bfs | BFS algorithm on the generated graph |
| local_clustering | Local clustering algorithm on the generated graph |
| redis_sets | 50M SET operations on redis |
| redis_gets | 50M GET operations on redis |
| memcached_sets | 50M SET operations on memcached |
| memcached_gets | 50M GET operations on memcached |

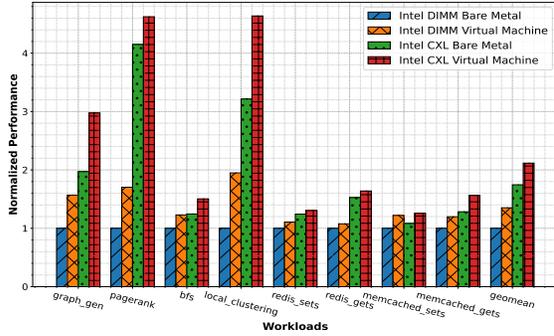
(a) Bare metal and VM normalized performance.

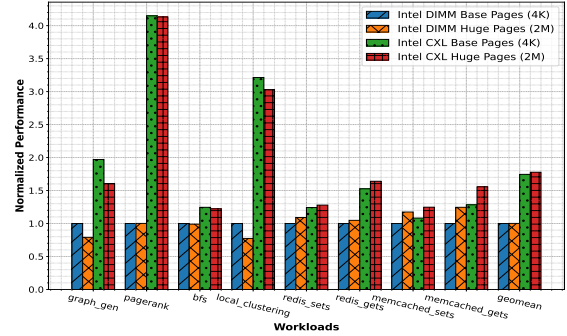
(b) 4KiB and 2MiB normalized performance.

Figure 28: Normalized performance for graph and KV workloads on Intel-DIMM and Intel-CXL.

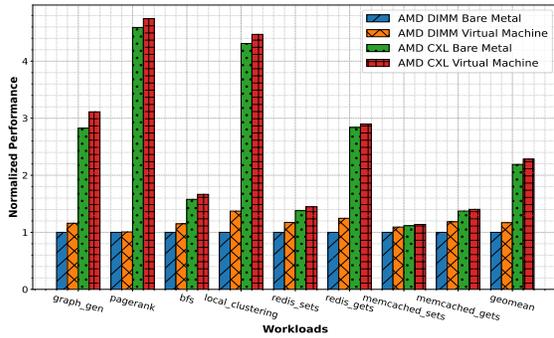
(a) Bare metal and VM normalized performance.

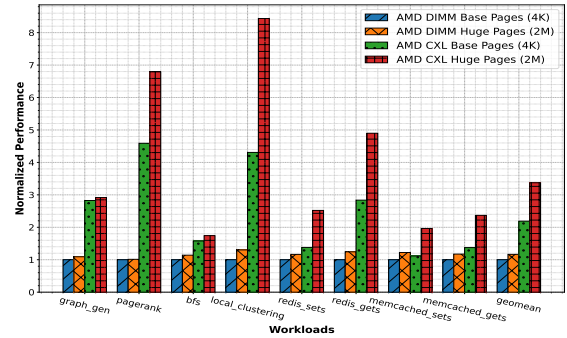
(b) 4KiB and 2MiB normalized performance.

Figure 29: Normalized performance for graph and KV workloads on AMD-DIMM and AMD-CXL.

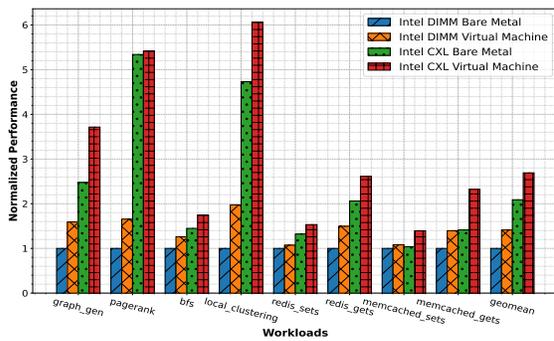
(a) Bare metal and VM normalized performance.

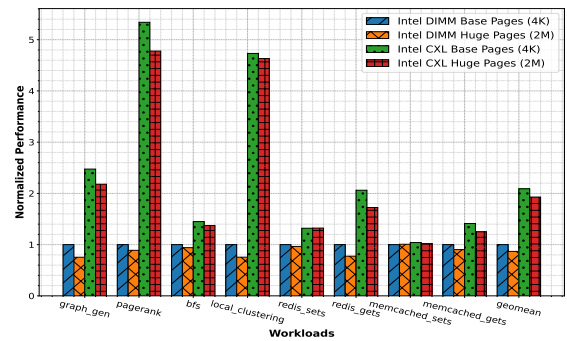
(b) 4KiB and 2MiB normalized performance.

Figure 30: Normalized performance for graph and KV workloads on Intel-DIMM and Intel-FPGA CXL.

# 7 Related Work

Several recent studies explored CXL-based systems' characteristics, performance, and programmability.

**Performance Characterization.** Sun et al. [20] evaluate several real and emulated CXL devices. They show that (a) real CXL devices can have better performance than NUMA-emulated CXL in some cases, (b) latency-sensitive applications suffer from CXL's higher access latency, and (c) propose a mechanism for dynamically tiering pages between DRAM and CXL devices. Liu et al. [37] compare and contrast CXL-attached memory devices with traditional DRAM-based NUMA node, explore use cases for CXL given its higher access latency compared to locally-attached DRAM, and go beyond Sun et al. and show that several HPC workloads can tolerate CXL-attached memory's high latency and low throughput. They also observed limited performance improvements in GPU LLM workloads when offloading tensors to CXL-attached memory. Tang et al. [38] evaluate the performance of CXL for cloud workloads and develop a model to estimate CXL's total cost of ownership (TCO). Previous works have also characterized other memory devices that offer an interface similar to CXL-attached memory [39–41].

**Memory Bandwidth and Capacity Expansion.** Pond [42] proposes a multi-host memory pooling solution to help with stranded memory in production cloud workloads. They show that a significant number of workloads show negligible performance loss when their entire memory is allocated in the memory pool. Levis et al. [43], however, argues against memory pooling for cloud workloads because of their cost, complexity, and utility. Yang et al. [44] evaluate a new DRAM-SSD hybrid CXL-based memory pooling architecture and show promising results for HPC workloads. Wahlgren et al. [45] developed a profiler for studying HPC workloads' memory access pattern and study capacity and bandwidth provisioning. Zhong et al. [46] developed a CXL memory allocator to intelligently allocate and place VM pages in multi-tenant hosts, minimizing workload slowdown. Transparent page placement (TPP) [11] is a mechanism to proactively place workload's cold pages in the far memory to enable faster near-memory allocations when new requests arrive. TPP can also dynamically and transparently move application pages between the near- and far-memory based on their *hotness*. TPP is submitted to be upstreamed into Linux [47]. Berger et al. [48] explore the design space of CXL-based memory pools and draw recommendations. They also study the cost savings from different pool sizes and find small pools to be the most cost-effective.

**Adapting applications for CXL.** Previous works have also looked at the performance of a number of applications on CXL. CC-NIC [49] implements a custom Host-NIC interface to take advantage of the CXL's byte-addressability and coherency, resulting in reduced latency and higher throughput compared to today's PCIe NICs. Other works [50–52] implement RPCs over CXL take advantage of the cache-coherent interface, improving performance. Many research works have also looked into CXL-enabled persistence [53–57]. Cabrera et al. [58] propose the use of CXL in heterogeneous accelerator systems to enable easier and fine-grained collaboration as opposed to host-enabled memory sharing.

Āpta [59] uses CXL-disaggregated memory to implement fault-tolerant object stores for function-as-a-service workloads outperforming RDMA-based solutions. Wei et al. [60] make a case for transaction index over CXL-based distributed memory using a lightweight distributed memory transaction primitive (rTX). rTX achieves significant performance improvement over other transactional indexes while minimizing the overhead of failure atomicity. Arelakis et al. [61] propose a CXL-attached memory expander that implements on-the-fly compression/decompression for hyperscalar workloads. They propose this as an alternative to software-based memory compression, saving CPU cycles.

# 8 Acknowledgement

The authors thank Hoshik Kim and Jongryool Kim for their valuable feedback. We also thank Kevin Xue for his contribution to this work. This paper is supported by the PRISM and ACE centers in JUMP 2.0, an SRC program sponsored by DARPA.

We acknowledge Giga Computing for providing access to and support for Gigabyte™ G383-R80 server equipped with 4 × AMD Instinct™ MI300A APUs.

# 9 Conclusion

In this paper, we develop a memory benchmark suite, HEIMDALL, and leverage it to study a wide range of CXL-attached systems' performance. Our observations span across the system architecture stack, from basic hardware behaviors, to micro-architecture and operating system performance, and applications including LLM inference and vector databases. These observations shed light on the future development of heterogeneous memory architecture and optimization in system software to better utilize such memory systems.

# References

[1] Denis Foley and John Danskin. "Ultra-Performance Pascal GPU and NVLink Interconnect". In: *IEEE Micro* (2017).

[2] Intel Corporation. *Intel® Xeon® Processor Scalable Family Technical Overview*. https://www.intel.com/content/www/us/en/developer/articles/technical/xeon-processor-scalable-family-technical-overview.html. 2022.

[3] AMD. *AMD Infinity Fabric*. https://en.wikichip.org/wiki/amd/infinity_fabric.

[4] CXL Consortium. *CXL Consortium releases Compute Express Link 3.0 specification*. URL: https://www.computeexpresslink.org/pressroom.

[5] Gen-Z Consortium. *The Gen-Z Consortium*. URL: https://genzconsortium.org/. (accessed: 06.24.2020).

[6] CCIX Consortium. *CCIX*. URL: https://www.ccixconsortium.com/. (accessed: 06.24.2020).

[7] Ying Wei, Yi Chieh Huang, Haiming Tang, Nithya Sankaran, Ish Chadha, Dai Dai, Olakanmi Oluwole, Vishnu Balan, and Edward Lee. "NVLink-C2C: A Coherent Off Package Chip-to-Chip Interconnect with 40Gbps/pin Single-ended Signaling". In: *2023 IEEE International Solid-State Circuits Conference (ISSCC)*. 2023.

[8] IBM. *Reaching the Summit: The World's Smartest Supercomputer*. https://newsroom.ibm.com/Reaching-the-Summit-The-Worlds-Smartest-Supercomputer. 2018.

[9] AMD. *AMD Instinct MI300A Accelerators*. https://www.amd.com/en/products/accelerators/instinct/mi300/mi300a.html.

[10] Ira Weiny. *cxl/pci: Skip irq features if MSI/MSI-X are not supported*. URL: https://lore.kernel.org/lkml/t6wkohozmtchuzzabjigr66tx6576nni54ig7lu2orlvqwmt5o@r52mxzei7uxs/T/.

[11] Hasan Al Maruf, Hao Wang, Abhishek Dhanotia, Johannes Weiner, Niket Agarwal, Pallab Bhattacharya, Chris Petersen, Mosharaf Chowdhury, Shobhit Kanaujia, and Prakash Chauhan. "TPP: Transparent Page Placement for CXL-Enabled Tiered-Memory". In: *Proceedings of the 28th ACM International Conference on Architectural Support for Programming Languages and Operating Systems*. ASPLOS 2023. 2023.

[12] Brendan Gregg. *Linux Perf Examples*. 2024. URL: http://www.brendangregg.com/perf.

[13] AMD. *AMD μProf*. 2024. URL: https://www.amd.com/en/developer/uprof.html.

[14] Intel. *Intel Performance Counter Monitor*. URL: https://github.com/intel/pcm.

[15] Timothy Prickett Morgan. *Intel 'Emerald Rapids' Xeon SPs: A Little More Bang, A Little Less Bucks*. 2023. URL: https://www.nextplatform.com/2023/12/14/intel-emerald-rapids-xeon-sps-a-little-more-bang-a-little-less-bucks/?t.

[16] NVIDIA. *Getting Started with Grace Hopper*. Accessed: 2025-04-08. NVIDIA Corporation. 2024. URL: https://docs.nvidia.com/clara/parabricks/latest/gettingstarted/gracehopper.html.

[17] Intel. *Intel 64 and IA-32 Architectures Optimization Reference Manual Volume 1*.

[18] Intel. *Intel Data Direct I/O Technology*. URL: https://www.intel.com/content/www/us/en/io/data-direct-i-o-technology.html.

[19] Alireza Farshin, Amir Roozbeh, Gerald Q Maguire Jr, and Dejan Kostić. "Reexamining Direct Cache Access to Optimize I/O Intensive Applications for Multi-hundred-gigabit Networks". In: *2020 USENIX Annual Technical Conference (USENIX ATC 20)*. 2020.

[20] Yan Sun, Yifan Yuan, Zeduo Yu, Reese Kuper, Chihun Song, Jinghan Huang, Houxiang Ji, Siddharth Agarwal, Jiaqi Lou, Ipoom Jeong, et al. "Demystifying CXL memory with genuine CXL-ready systems and devices". In: *Proceedings of the 56th Annual IEEE/ACM International Symposium on Microarchitecture (MICRO)*. 2023.

[21] Intel. *Introduction to Cache Allocation Technology in the Intel Xeon Processor E5 v4 Family*. URL: https://www.intel.com/content/www/us/en/developer/articles/technical/introduction-to-cache-allocation-technology.html.

[22] Intel. *Intel 64 and IA-32 Architectures Software Developer Manuals*. URL: https://www.intel.com/content/www/us/en/developer/articles/technical/intel-sdm.html.

[23] AMD. *AMD64 Architecture Programmer's Manual, Volumes 1-5*. URL: https://www.amd.com/en/support/tech-docs.

[24] Intel. *Intel Memory Latency Checker v3.11a*. URL: https://www.intel.com/content/www/us/en/developer/articles/tool/intelr-memory-latency-checker.html.

[25] Jason Ansel, Edward Yang, Horace He, Natalia Gimelshein, Animesh Jain, Michael Voznesensky, Bin Bao, Peter Bell, David Berard, Evgeni Burovski, Geeta Chauhan, Anjali Chourdia, Will Constable, Alban Desmaison, Zachary DeVito, Elias Ellison, Will Feng, Jiong Gong, Michael Gschwind, Brian Hirsh, Sherlock Huang, Kshiteej Kalambarkar, Laurent Kirsch, Michael Lazos, Mario Lezcano, Yanbo Liang, Jason Liang, Yinghai Lu, C. K. Luk, Bert Maher, Yunjie Pan, Christian Puhrsch, Matthias Reso, Mark Saroufim, Marcos Yukio Siraichi, Helen Suk, Shunting Zhang, Michael Suo, Phil Tillet, Xu Zhao, Eikan Wang, Keren Zhou, Richard Zou, Xiaodong Wang, Ajit Mathews, William Wen, Gregory Chanan, Peng Wu, and Soumith Chintala. "PyTorch 2: Faster Machine Learning Through Dynamic Python Bytecode Transformation and Graph Compilation". In: *Proceedings of the 29th ACM International Conference on Architectural Support for Programming Languages and Operating Systems, Volume 2*. ASPLOS '24. 2024.

[26] Georgi Gerganov. *LLama.cpp: Port of Meta's LLaMA model to C/C++*. 2023. URL: https://github.com/ggerganov/llama.cpp.

[27] Woosuk Kwon, Zhuohan Li, Siyuan Zhuang, Ying Sheng, Lianmin Zheng, Cody Hao Yu, Joseph E. Gonzalez, Hao Zhang, and Ion Stoica. "Efficient Memory Management for Large Language Model Serving with PagedAttention". In: *Proceedings of the ACM SIGOPS 29th Symposium on Operating Systems Principles*. 2023.

[28] Stephen Merity, Caiming Xiong, James Bradbury, and Richard Socher. *Pointer Sentinel Mixture Models*. 2016.

[29] OpenChat. *OpenChat ShareGPT V3 Dataset*. 2024. URL: https://huggingface.co/datasets/openchat/openchat_sharegpt_v3.

[30] Hugo Touvron, Thibaut Lavril, Gautier Izacard, Xavier Martinet, Marie-Anne Lachaux, Timothée Lacroix, Baptiste Rozière, Naman Goyal, Eric Hambro, Faisal Azhar, Aurelien Rodriguez, Armand Joulin, Edouard Grave, and Guillaume Lample. "LLaMA: Open and Efficient Foundation Language Models". In: (2023). URL: https://arxiv.org/abs/2302.13971.

[31] Qdrant. *Qdrant: High-Performance Vector Search at Scale*. URL: https://qdrant.tech/.

[32] milvus. *Milvus: The High-Performance Vector Database Built for Scale*. URL: https://milvus.io/.

[33] Tensorflow. *glove100 angular dataset*. URL: https://www.tensorflow.org/datasets/catalog/glove100_angular.

[34] QEMU Project. *QEMU/KVM*. URL: https://www.qemu.org/.

[35] Redis Inc. *redis-py - Python Client for Redis*. URL: https://redis-py.readthedocs.io/en/stable/.

[36] Sean Reifschneider. *python-memcached*. URL: https://github.com/linsomniac/python-memcached.

[37] Jie Liu, Xi Wang, Jianbo Wu, Shuangyan Yang, Jie Ren, Bhanu Shankar, and Dong Li. *Exploring and Evaluating Real-world CXL: Use Cases and System Adoption*. 2024. URL: https://arxiv.org/abs/2405.14209.

[38] Yupeng Tang, Ping Zhou, Wenhui Zhang, Henry Hu, Qirui Yang, Hao Xiang, Tongping Liu, Jiaxin Shan, Ruoyun Huang, Cheng Zhao, Cheng Chen, Hui Zhang, Fei Liu, Shuai Zhang, Xiaoning Ding, and Jianjun Chen. "Exploring Performance and Cost Optimization with ASIC-Based CXL Memory". In: *Proceedings of the Nineteenth European Conference on Computer Systems*. EuroSys '24. 2024.

[39] Onkar Patil, Latchesar Ionkov, Jason Lee, Frank Mueller, and Michael Lang. "Performance characterization of a DRAM-NVM hybrid memory architecture for HPC applications using intel optane DC persistent memory modules". In: *Proceedings of the International Symposium on Memory Systems*. MEMSYS '19. 2019.

[40] Zixuan Wang, Xiao Liu, Jian Yang, Theodore Michailidis, Steven Swanson, and Jishen Zhao. "Characterizing and Modeling Non-Volatile Memory Systems". In: *2020 53rd Annual IEEE/ACM International Symposium on Microarchitecture (MICRO)*. 2020.

[41] Joseph Izraelevitz, Jian Yang, Lu Zhang, Juno Kim, Xiao Liu, Amirsaman Memaripour, Yun Joon Soh, Zixuan Wang, Yi Xu, Subramanya R. Dulloor, Jishen Zhao, and Steven Swanson. *Basic Performance Measurements of the Intel Optane DC Persistent Memory Module*. 2019. URL: https://arxiv.org/abs/1903.05714.

[42] Huaicheng Li, Daniel S. Berger, Lisa Hsu, Daniel Ernst, Pantea Zardoshti, Stanko Novakovic, Monish Shah, Samir Rajadnya, Scott Lee, Ishwar Agarwal, Mark D. Hill, Marcus Fontoura, and Ricardo Bianchini. "Pond: CXL-Based Memory Pooling Systems for Cloud Platforms". In: *Proceedings of the 28th ACM International Conference on Architectural Support for Programming Languages and Operating Systems, Volume 2*. ASPLOS 2023. 2023.

[43] Philip Levis, Kun Lin, and Amy Tai. "A Case Against CXL Memory Pooling". In: *Proceedings of the 22nd ACM Workshop on Hot Topics in Networks*. HotNets '23. 2023.

[44] Qirui Yang, Runyu Jin, Bridget Davis, Devasena Inupakutika, and Ming Zhao. "Performance Evaluation on CXL-enabled Hybrid Memory Pool". In: *2022 IEEE International Conference on Networking, Architecture and Storage (NAS)*. 2022.

[45] Jacob Wahlgren, Maya Gokhale, and Ivy B. Peng. "Evaluating Emerging CXL-enabled Memory Pooling for HPC Systems". In: *2022 IEEE/ACM Workshop on Memory Centric High Performance Computing (MCHPC)*. 2022.

[46] Yuhong Zhong, Daniel S. Berger, Carl Waldspurger, Ryan Wee, Ishwar Agarwal, Rajat Agarwal, Frank Hady, Karthik Kumar, Mark D. Hill, Mosharaf Chowdhury, and Asaf Cidon. "Managing Memory Tiers with CXL in Virtualized Environments". In: *18th USENIX Symposium on Operating Systems Design and Implementation (OSDI 24)*. 2024. URL: https://www.usenix.org/conference/osdi24/presentation/zhong-yuhong.

[47] Hasan Al Maruf. *Transparent Page Placement for Tiered-Memory*. https://lore.kernel.org/all/cover.1637778851.git.hasanalmaruf@fb.com/. 2021.

[48] Daniel S. Berger, Daniel Ernst, Huaicheng Li, Pantea Zardoshti, Monish Shah, Samir Rajadnya, Scott Lee, Lisa Hsu, Ishwar Agarwal, Mark D. Hill, and Ricardo Bianchini. "Design Tradeoffs in CXL-Based Memory Pools for Public Cloud Platforms". In: *IEEE Micro* (2023).

[49] Henry N Schuh, Arvind Krishnamurthy, David Culler, Henry M Levy, Luigi Rizzo, Samira Khan, and Brent E Stephens. "CC-NIC: a Cache-Coherent Interface to the NIC". In: *Proceedings of the 29th ACM International Conference on Architectural Support for Programming Languages and Operating Systems, Volume 1*. 2024.

[50] Mingxing Zhang, Teng Ma, Jinqi Hua, Zheng Liu, Kang Chen, Ning Ding, Fan Du, Jinlei Jiang, Tao Ma, and Yongwei Wu. "Partial Failure Resilient Memory Management System for (CXL-based) Distributed Shared Memory". In: *Proceedings of the 29th Symposium on Operating Systems Principles*. SOSP '23. 2023.

[51] Suyash Mahar, Ehsan Hajyjasini, Seungjin Lee, Zifeng Zhang, Mingyao Shen, and Steven Swanson. *Telepathic Datacenters: Fast RPCs using Shared CXL Memory*. 2024. URL: https://arxiv.org/abs/2408.11325.

[52] Jie Zhang, Xuzheng Chen, Yin Zhang, and Zeke Wang. "DmRPC: Disaggregated Memory-aware Datacenter RPC for Data-intensive Applications". In: *2024 IEEE 40th International Conference on Data Engineering (ICDE)*. 2024.

[53] Yehonatan Fridman, Suprasad Mutalik Desai, Navneet Singh, Thomas Willhalm, and Gal Oren. *CXL Memory as Persistent Memory for Disaggregated HPC: A Practical Approach*. 2023. URL: https://arxiv.org/abs/2308.10714.

[54] Sangjin Lee, Alberto Lerner, Philippe Bonnet, and Philippe Cudré-Mauroux. "Database Kernels: Seamless Integration of Database Systems and Fast Storage via CXL." In: *CIDR*. 2024.

[55] Myoungsoo Jung. "Hello bytes, bye blocks: Pcie storage meets compute express link for memory expansion (cxl-ssd)". In: *Proceedings of the 14th ACM Workshop on Hot Topics in Storage and File Systems*. 2022.

[56] Suyash Mahar, Mingyao Shen, Terence Kelly, and Steven Swanson. "Snapshot: Fast, Userspace Crash Consistency for CXL and PM Using msync". In: *2023 IEEE 41st International Conference on Computer Design (ICCD)*. 2023.

[57] Yi Xu, Suyash Mahar, Ziheng Liu, Mingyao Shen, and Steven Swanson. *CXL Shared Memory Programming: Barely Distributed and Almost Persistent*. 2024. URL: https://arxiv.org/abs/2405.19626.

[58] Anthony M Cabrera, Aaron R Young, and Jeffrey S Vetter. "Design and analysis of CXL performance models for tightly-coupled heterogeneous computing". In: *Proceedings of the 1st International Workshop on Extreme Heterogeneity Solutions*. ExHET '22. 2022.

[59] Adarsh Patil, Vijay Nagarajan, Nikos Nikoleris, and Nicolai Oswald. "Āpta: Fault-tolerant object-granular CXL disaggregated memory for accelerating FaaS". In: *2023 53rd Annual IEEE/IFIP International Conference on Dependable Systems and Networks (DSN)*. 2023.

[60] Xingda Wei, Haotian Wang, Tianxia Wang, Rong Chen, Jinyu Gu, Pengfei Zuo, and Haibo Chen. *Transactional Indexes on (RDMA or CXL-based) Disaggregated Memory with Repairable Transaction*. 2023. URL: https://arxiv.org/abs/2308.02501.

[61] Angelos Arelakis, Nilesh Shah, Yiannis Nikolakopoulos, and Dimitrios Palyvos-Giannas. *Streamlining CXL Adoption for Hyperscale Efficiency*. 2024. URL: https://arxiv.org/abs/2404.03551.